\documentclass[acmlarge,nonacm]{acmart}

\AtBeginDocument{%
  \providecommand\BibTeX{{%
    \normalfont B\kern-0.5em{\scshape i\kern-0.25em b}\kern-0.8em\TeX}}}

\setcopyright{cc}
\setcctype{by-sa}
\copyrightyear{2026}
\acmYear{2026}

\acmConference[Conference acronym 'XX]{Make sure to enter the correct
  conference title from your rights confirmation emai}{June 03--05,
  2018}{Woodstock, NY}
\acmISBN{978-1-4503-XXXX-X/18/06}

\usepackage{enumitem}
\usepackage{quoting}
\quotingsetup{font=itshape,vskip=1pt, leftmargin=15pt}
\usepackage{cleveref}
\usepackage{graphicx}
\usepackage{subcaption}
\usepackage{tikz}
\usetikzlibrary{arrows, arrows.meta, decorations.pathreplacing,	calligraphy, positioning, fit, calc, backgrounds, decorations.text, external}
\usepackage{pgfplots}
\usepackage{pgfplotstable}
\pgfplotsset{compat=1.18}

\definecolor{Peach}{RGB}{246, 199, 175}
\definecolor{SurgicalGreen}{RGB}{0, 123, 116}
\definecolor{boxGray}{RGB}{199, 199, 199}
\definecolor{brightPink}{RGB}{255,0,244}

\definecolor{shadecolor}{rgb}{0.96, 0.80, 0.71}

\newcommand{\old}[1]{\unskip}

\newcommand{\prototype}{\emph{SkeletonDance}}
\newcommand{\GMPs}{\textit{Generalized Motor Programs (GMPs)}}
\newcommand{\GMP}{\textit{Generalized Motor Program (GMP)}}
\newcommand{\distancemetric}{$\Delta_{leg}$}

\newcommand{\beatshit}{rhythmic accuracy}
\newcommand{\amtlost}{rhythm lost count}
\newcommand{\meanofftimes}{mean out-of-rhythm duration}
\newcommand{\fbduration}{feedback duration}

\newcommand{\Beatshit}{Rhythmic Accuracy}
\newcommand{\Amtlost}{Rhythm Lost Count}
\newcommand{\Meanofftimes}{Mean Out-Of-Rhythm Duration}
\newcommand{\Fbduration}{Feedback Duration}

\newcommand{\samplemean}{\bar{x}}
\newcommand{\samplesd}{s}
\newcommand{\stderror}{SE}

\newcommand{\fb}{\textsc{feedback}}
\newcommand{\base}{\textsc{baseline}}

\makeatletter
\newcommand\footnoteref[1]{\protected@xdef\@thefnmark{\ref{#1}}\@footnotemark}
\makeatother

\begin{document}




\title[Designing Interactive Rhythm Feedback]{Rhythm Is a Dancer: Designing Interactive Rhythm Feedback for Beginner Dancers}
\author{Bettina Eska}
\orcid{0000-0002-0954-1394}
\affiliation{%
  \institution{LMU Munich}
  \city{Munich}
  \country{Germany}
}
\email{bettina.eska@ifi.lmu.de}

\author{Annika Kilian}
\orcid{0000-0002-1407-7303}
\affiliation{%
  \institution{LMU Munich}
  \city{Munich}
  \country{Germany}
}
\email{annikaxkilian@gmail.com}

\author{Pawe\l{}~W.~Wo\'zniak}
\orcid{0000-0003-3670-1813}
\affiliation{%
  \institution{TU Wien} 
  \city{Vienna}
  \country{Austria}
}
\email{pawel.wozniak@tuwien.ac.at}

\author{Jakob Karolus}
\email{jakob.karolus@dfki.de}
\orcid{0000-0002-0698-7470}
\affiliation{%
	\institution{German Research Center for Artificial Intelligence (DFKI)}
	\city{Kaiserslautern}
	\country{Germany}
}
\affiliation{%
	\institution{RPTU Kaiserslautern-Landau}
    \city{Kaiserslautern}
	\country{Germany}
}
\renewcommand{\shortauthors}{Eska, et al.}

\begin{abstract}

Learning how to dance can readily overwhelm beginners, especially without effective guidance from a dance teacher. 
Existing interactive systems often do not sufficiently support the learner’s progress.
We investigated how targeted feedback on rhythm keeping interactively supports dance practice for novice dancers by introducing \prototype{}.
Our design is grounded in motor learning theory and conceptualized through interviews with dance teachers, following established teaching strategies. \prototype{} automatically detects rhythm flaws and provides assistance through mimicking clapping feedback, a common instructional technique in dance lessons. 
In our study, participants reported that \prototype{} helped them to re-establish lost rhythm and increased confidence during practice, especially among novices.
Though objective performance metrics did not consistently confirm these effects during controlled test sessions. 
Our work highlights that feedback can support novice dancers’ subjective practicing experiences and demonstrates how prior dancing experience moderates the objective effectiveness of such minimal, teacher-inspired interventions.

\end{abstract}


\begin{teaserfigure}
\centering
  \includegraphics[width=.4\textwidth]{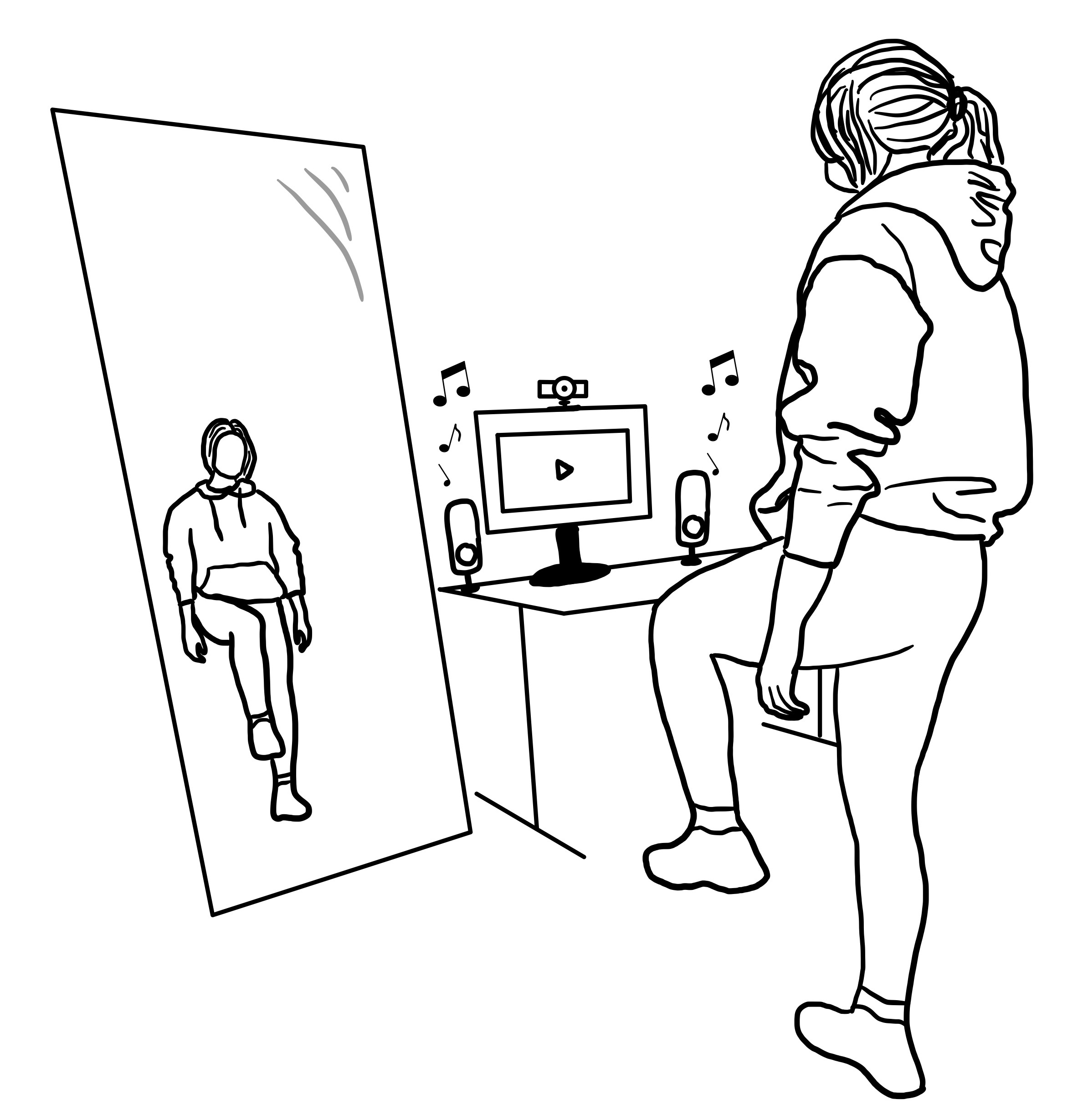}
  \caption{\prototype{} supports users in practicing dance sequences. The system provides auditory feedback on rhythm accuracy which is detected through key body movements.}
  \label{fig:teaser}
  \Description{User learning a new dance sequence. The woman is standing in front of a mirror and lifts her left leg. Next to the right of the mirror is a table with a monitor, a webcam on top, and two speakers. Music notes around the speakers symbolize that music is playing. }
\end{teaserfigure}

\maketitle

\section{Introduction}
\label{sec:intro}
Dancing is an integral part of different cultures, can define identity, and constitutes part of our daily lives~\cite{blacking1983movementmeaning}. Entertainment to a musical beat is part of human nature~\cite{lalandEvolutionDance2016}. Yet, despite the fact that dancing has gained popularity with the help of games, apps, and social media platforms in the last few years, developing dance skills comes with significant challenges. The coordinated motion involved in dance is a complex skill that requires extensive training and physiological development for most users~\cite{malkogeorgos2013physiological}. As a complex athletic pursuit, dancing can offer well-being benefits both in terms of stress reduction and health benefits stemming from physical activity~\cite{tao2022physiological}.

The high complexity makes dance learning a fertile ground for interactive technology to support the process~\cite{zhou2021syncupvisionbased,trajkovaDesigningBalletClasses2019}. Thus, it is a challenge for Human-Computer Interaction (HCI) to understand how to design systems that effectively support learning to dance.

Research has already explored systems that support timing, performance, and movement execution~\cite{rahebDanceInteractiveLearning2019, diaspereiradossantosLetDanceHow2017, andersonYouMoveEnhancingMovement2013, grosshauserWearableSensorBased2011, turmovidalEnlightenedYogaDesigning2019}. State-of-the-art dance games, such as 
\textit{Dance Dance Revolution (DDR)}~\cite{konamiDanceDanceRevolution2013}, \textit{Just Dance}~\cite{ubisoftJustDanceUnlimited2021}, and \textit{Dance Central}~\cite{mtvgamesDanceCentral2010} have enjoyed increased popularity. Yet, past solutions are mainly score-based, such as awarding points on how well the user imitates target movements. This approach --- while often effective in the social context of a party game --- has limited applications for those who want to perceive dancing as a hobby or outside of competition. Due to the complexity of dancing, this score-based feedback does not allow the user to gain a deeper understanding of dancing. Consequently, informed reasoning about their mistakes and derived improvements are more difficult to achieve.

Research has explored feedback systems for dancing that address the quality of the movement~\cite{grosshauserWearableSensorBased2011} and the posture~\cite{maharaj-pariagsinghDanceTutorITSCoaching2021} and support the user with feedback on the rhythm~\cite{diaspereiradossantosLetDanceHow2017}. However, the systems in related works often require complex setups, such as full-body motion tracking systems, which makes them impractical for everyday use. 
Thus, our work targets alternative feedback forms that work with minimal setup and facilitate initial engagement for novice dancers.


We present an inquiry where we explore how interactive technology can support the complexity of practicing dance. To build an understanding of the design space, we draw from existing methodologies in literature~\cite{lukoffAncientContemplativePractice2020, trajkovaDesigningBalletClasses2019} conducting interviews with dance teachers. We identified common teaching approaches and student problems in dance practice. We further linked those to concepts from motor learning theory. One of the most common difficulties mentioned by the dance teachers is maintaining rhythm and timing. 
Based on these findings, we developed and implemented \prototype{}, an interactive feedback system to support beginner dancers with practicing dance in rhythm. The system uses a camera to record the dancer and to generate unobtrusive but noticeable auditory feedback while the person is practicing. When the dancer is not in sync with the rhythm of the music, the system plays a clapping sound as an overlay to the music to enhance the beat, mimicking the instructional technique commonly used by dance teachers in this situation.

In an evaluation, we used \prototype{} to mimic a real dance lesson. Novice dancers were asked to practice two dance sequences in practice sessions using video instructions, once being aided by the clapping feedback provided by \prototype{}. Afterwards, we evaluated their dance performance by measuring rhythmic accuracy in test sessions without any support, and asked participants about their subjective practicing experiences.
Although objective metrics of rhythm accuracy did not show a significant effect of the feedback condition, potentially due to differences in prior dancing experience, participants reported that \prototype{} provided noticeable feedback that helped them regain rhythm and feel more secure, easing access to dance practice.

This paper contributes the following: (1) an investigation into applicable dance teaching techniques for interactive systems grounded in motor learning theory and interviews with dance instructors; (2) the resulting design and evaluation of \prototype{}---an interactive system that supports dance practicing through rhythm-based detection and feedback; and (3) insights into designing technologies that support dance practice.

\section{Related Work}
\label{sec:relatedwork}



Dancing is a creative process, as such, it is inherently difficult which features are most suited to capture the expressive qualities of dancing~\cite{zhouDanceChoreographyHCI2021}. The complexity of the movements coupled with the dancer's personal interpretation of music and intertwined movements poses significant challenges for any interactive support system. In this section, we present existing research works in this domain and how they address this challenge. 

\subsection{Interactive Dance Systems and Feedback}\label{subsec:ids_fb} 
As dancing is an important aspect of many cultures and a popular leisure activity for many people, there are many opportunities to engage in it. 
Dance learning is supported both by commercial systems and research prototypes, which differ in their feedback strategies and learning goals.

\subsubsection{Dance Games} Commercial dance games are usually referred to as esports or aerobic training with an entertainment focus.
Famous examples of dance games are, e.g., the arcade game \textit{Dance Dance Revolution (DDR)}~\cite{konamiDanceDanceRevolution2013} or the motion-based games \textit{Just Dance}~\cite{ubisoftJustDanceUnlimited2021} and \textit{Dance Central}~\cite{mtvgamesDanceCentral2010}. 
These games focus in entertainment by instructing users to follow the movement by tapping shown arrows on the dance pad in rhythm with the music (DDR) or imitate the choreography on the screen.
These systems evaluate timing and movement accuracy via input devices (e.g., dance pads, controllers, cameras) and provide score-based feedback (e.g., `Perfect', `Good', `Miss'). 
However, this feedback remains coarse-grained and does not explain errors or support reflection. Users must infer whether mistakes stem from timing or movement inaccuracies. 
In contrast, our work focuses on supporting the practicing progress through feedback, which encourages reflection by the student to foster the practicing of a new skill.

\subsubsection{Dance Education Systems}
In contrast to dance games, HCI research has also looked into dance education systems while focusing more on the correct technique and feedback mechanisms tailored to dance performance.
Approaches include gamified feedback (e.g., color-coded positions)~\cite{charbonneauTeachMeDance2011} and movement execution evaluations~\cite{grosshauserWearableSensorBased2011}. 
\citet{rahebDanceInteractiveLearning2019} reviewed other systems and identified challenges in the field of dance teaching and learning, which include genre, learning approaches, context, and learner profile. 
They use different communication modes and types of feedback, depending on the teaching goal. To effectively support motor learning, experts usually provide augmented feedback~\cite{sigristAugmentedVisualAuditory2013} on small chunks to support learning~\cite{krasnowMotorLearningControl2015, lindsayAdaptableCoachCritical2024}.
In practice, teachers demonstrate movements, explain them verbally or through gestures, and provide corrections through touch~\cite{gibbonsTeachingDanceSpectrum2007}.



Mirrors are a common tool used for immediate visual feedback through self-observation, improving performance during early learning stages despite concerns about body image~\cite{radellImpactMirrorsBody2004, trajkovaDesigningBalletClasses2019, dearborn2006dancelearning}.

\subsection{Sensing Technology and Dancing}
Consequently, enriching training through sensor-based detection systems, like motion-tracking~\cite{trajkovaDesigningBalletClasses2019}, is beneficial, yet rises the challenge on how feedback can be delivered. 
Emerging sensing technologies leverage motion tracking with a wide range of systems, such as full-body motion tracking, Inertial Measurement Units (IMUs), wearable sensors, virtual reality (VR), augmented reality (AR), or mixed reality (MR), and smartphone applications~\cite{santosTrainingBodyPotential2016, zhouMovementGuidanceUsing2022}. They often generate feedback using the deviation of the learner's sensor data from the performance of an expert.

However, there is still a large range in how to design feedback to enhance practice. Feedback varies from self-reflection support~\cite{eska2023hiit} to explicit annotations and demonstrations using videos or avatars~\cite{rahebDanceInteractiveLearning2019}. 
Other setups rely on vibrotactile actuators in addition to visual demonstrations~\cite{camarillo-abadEnvironmentMotorSkill2021, drobnyLearningBasicDance2010, nakamuraMultimodalPresentationMethod2005}.

Two big branches in the research about dance supporting systems are dance movement technique and rhythm. 
Systems focus on improving execution quality. For example, \textit{YouMove}~\cite{andersonYouMoveEnhancingMovement2013} uses an AR mirror with graphical overlays and specified crucial joints at certain moments, which later appeared as the feedback. 
%
In this work, we opted for a camera-based approach, allowing easy set-up and deployment.

On the rhythm side, other systems focus on assessing the synchronism with the beat of the music~\cite{diaspereiradossantosLetDanceHow2017, drobnyLearningBasicDance2010, bergner2019firststepsdance} and group coordination~\cite{zhou2021syncupvisionbased}. 
Applications such as the Forró Trainer extract features, such as rhythm duration, consistency, and body motion from the motion data, and provide retrospective feedback~\cite{diaspereiradossantosLetDanceHow2017}, while accelerometer-based approaches evaluate rhythm against expert assessments and provide performance analysis with summaries, visualizations, and narratives.~\cite{diaspereiradossantosYouAreBeat2018}. 
Other systems provide real-time feedback, e.g., emphasizing beats when timing is incorrect~\cite{drobnyLearningBasicDance2010}, though sensing limitations can reduce accuracy.


\subsection{Towards a Deeper Understanding of Dancing and Bodily Experiences}
There are plentiful technologies available to track and evaluate dance performance as outlined above. However, a question that remains is how to design feedback systems effectively to support the body movements. Finding the right balance between providing comprehensive insights into performance while ensuring the feedback is clear and encourages constructive self-reflection of bodily experiences is still challenging~\cite{ley-flores2024codesigningsensory}. Moreover, feedback should not only aim to enhance the technical proficiency of the dance moves. Still, it should also generate proprioception and a deeper understanding of one's own body~\cite{pylvanainen2003body} in the context of dance. Eliciting focused body sensations can help overcome barriers to physical movements~\cite{ley-flores2024codesigningsensory}. Consequently, providing the right stimuli at the right time --- like imitating a dance teacher --- is one of the main design goal of \prototype.
\section{Methodology: From Theory to Practice}
\label{sec:method}
In this work, we aim to investigate how interactive feedback designs can support beginner dancers during practice. While motor learning theory provides a conceptual lens for when and how feedback might support dance practice, our work does not aim to empirically evaluate or validate motor learning theories themselves. Evaluating motor learning, particularly in complex activities such as dance, is methodologically challenging, as learning involves gradual and often implicit changes in movement execution, proprioception, and bodily awareness that are especially difficult to capture through quantitative metrics in a short-time evaluation.

Rather than operationalizing motor learning as a measurable outcome, we therefore primarily use theory to guide design decisions and focus our evaluation on dancers’ subjective experiences. This aligns with our primary goal to assess whether the feedback design of \prototype{} offers meaningful support from the user’s perspective. Accordingly, our methodology emphasizes participants’ subjective experiences using questionnaires and semi-structured interviews. Formally, our research is governed by the following research questions:

\begin{enumerate}
\item[\textbf{RQ1:}] What are design requirements of an interactive feedback system for beginner dancers?
    \begin{description}[leftmargin=1em, labelindent=-1em]
    \item[\textit{\textbf{Theory:}}]
    \label{RQ1a}
        Which theories from motor learning can inform the design of feedback system for physical movements, in particular for dance movements?
        
        \textit{Objective:} Identifying relevant concepts from theory. (1) When to deliver feedback, (2) What feedback to deliver and (3) How.
        
        \textit{Method:} Review of seminal works in the domain of motor learning theory in the context of dance movements.
    
    \item[\textit{\textbf{Practice:}}]
    \label{RQ1b}
        Which teaching methods have empirically been proven to be successful and how can they be integrated in interactive feedback systems?
        
        \textit{Objective:} Gather insights from dance teachers and identify key difficulties for beginner dancers and how teachers address them in regular dance classes.
    
        \textit{Method:} Collect insights from interviews with experienced dance teachers and consolidate them with relevant concepts from motor learning theories.
    \end{description}

\item[\textbf{RQ2:}] Does \prototype{} support beginner dancers during practice, particularly with respect to rhythm keeping?
    \begin{description}[leftmargin=1em, labelindent=-1em]
    \item[\textit{\textbf{Rhythm Performance:}}]
    \label{RQ2a}
       Do dancers achieve a higher ratio of in-sync dancing when training with \prototype?
        
        \textit{Objective:} Evaluate if \prototype{} allows users to keep to the rhythm more often.
        
        \textit{Method:} Quantitative analysis of rhythmic accuracy based on body movements, and assessment of self-perceived rhythmic accuracy through questionnaires.
    
    \item[\textit{\textbf{Body Perception:}}]
        \label{RQ2b}
        Do dancers achieve a better understanding of their own body movements when training with \prototype?
    
        \textit{Objective:} Gather qualitative insights into how \prototype{} affects the body perception of the dancers and whether it supports increased body awareness.

        \textit{Method:} Thematic analysis of post-study interviews, and questionnaires on self-perceived body awareness and perception.
    \end{description}
\end{enumerate}

\begin{figure*}[h]
	\centering
	\begin{tikzpicture}
		
		\node at (0,0) (interviews) [text width=3.8cm, draw, rounded corners, text centered, thick, minimum height=1cm, minimum width=4cm] {\parbox{3.5cm}{\centering\textbf{Dance Teacher\\ Interviews (\Cref{sec:e_interviews})}}};
		\node at (0,-3) (theory) [text width=3.8cm, draw, rounded corners, text centered, thick, minimum height=1cm, minimum width=4cm] {\parbox{3.5cm}{\centering\textbf{Motor Learning\\ Theory (\Cref{subsec:theory})}}};
		\node at (5,-1.5) (design) [text width=3.8cm, draw, rounded corners, text centered, thick, minimum height=1.5cm, minimum width=4cm] {\parbox{3.5cm}{\centering\textbf{Design and Implementation of\\ \prototype{} (\Cref{sec:designimpl})}}};
		\node at (10,-1.5) (eval) [text width=3.8cm, draw, rounded corners, text centered, thick, minimum height=1.5cm, minimum width=4cm] {\parbox{3.5cm}{\centering\textbf{Evaluation of\\ \prototype{} (\Cref{sec:study})}}};
		
        \draw[-{Latex[length=5mm, width=2mm]}, thick] (interviews.south) |- ($(design.south west)!0.66!(design.north west)$);
        \draw[-{Latex[length=5mm, width=2mm]}, thick] (theory.north) |- ($(design.south west)!0.33!(design.north west)$);
		\draw[-{Latex[length=5mm, width=2mm]}, thick] (design.east) to (eval.west);

	\end{tikzpicture}
	\caption{Overview of our methodology. We combined background knowledge from motor learning theories and insights into empirical teaching leveraged from interviews with dance teachers to design \prototype{}.}
	\label{fig:method}
        \Description{Flow graph of the procedure in our investigation. Consisting of a field for Dance Teacher Interviews (Section 5) and below a field with Motor Control Theories (Section 4). Both fields feed into the Design and Implementation of SkeletonDance (Section 6). At the end stands the Evaluation of SkeletonDance (Section 7).}
\end{figure*}
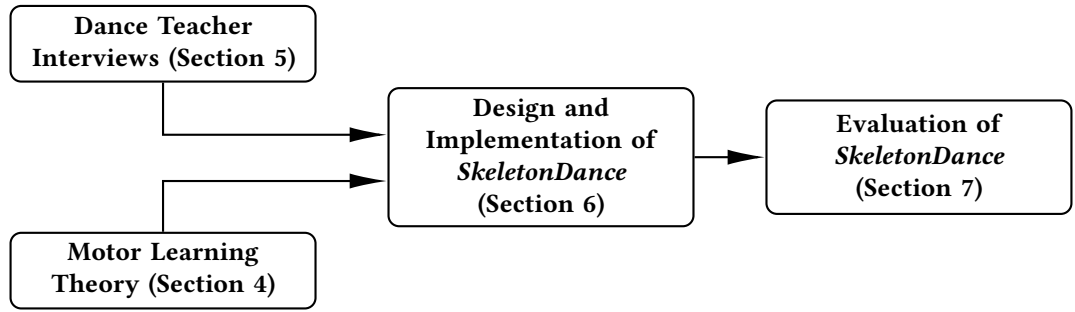

A procedural overview of our methodology is shown in \Cref{fig:method}. We first review established theories of motor control learning and highlight relevant concepts to our work (\Cref{subsec:theory}). To find out how motor learning theories are applied in practice, we then conducted interviews with experienced dance teachers (\Cref{sec:e_interviews}). They often use teaching methods that are well-proven over time.
The interviews yielded helpful domain knowledge and provided insights into tools and feedback currently used during dance classes.

We subsequently linked the insights from the interviews with concepts from motor learning theories and derived design requirements for \prototype{} (\Cref{subsec:designPT}).
These outcomes were then incorporated into the design and implementation of \prototype. We further considered the requirement that \prototype{} should be capable of being integrated into existing dance practice and readily deployable to enable easy access by everyone.
We then evaluated our system in a user study (\Cref{sec:study}) measuring rhythmic accuracy of participants. We also added questionnaires and post-hoc interviews targeting the provided feedback and its effect on the users’ proprioception. We compared these results within each participant between the baseline (video only) and feedback (video and clapping feedback) condition.

\section{Theory: Motor Learning and Dancing} 
\label{subsec:theory}
To understand how dance movements are learned and consolidated, we draw on established theories of motor learning, which inform the design our work.

\subsection{Phases of Motor Learning}
\label{subsubsec:phasesML}
Motor learning~\cite{schmidtMotorControlLearning2018} can be defined as ``a set of processes associated with practice or experience leading to relatively permanent changes in the capability for movement''~\cite{zwickerReflectionMotorLearning2009}. 
Throughout the learning process, a motor skill is developed in different cognitive stages~\cite{fittsHumanPerformance1967}: \textit{cognitive, associative,} and \textit{autonomous}. The skill evolves from a general idea about the movement or task with more errors (cognitive) over the refinement of the skill through practice (associative) and learning from errors to the stage when the skill becomes automatic without requiring a lot of cognitive effort~\cite{zwickerReflectionMotorLearning2009}. In this process, apart from practice, feedback is one of the most critical components of improving a skill~\cite{sharmaEffectivenessKnowledgeResult2016, winsteinKnowledgeResultsMotor1991, rahebDanceInteractiveLearning2019}. 


\subsection{The Schema Theory - Generalized Motor Programs}
\label{subsubsec:schematheory}
While the allocation of motor learning into phases offers a temporal view at motor skill development, the \textit{schema theory} by \citet{schmidtMotorControlLearning2018} 
provides a cross-sectional perspective by introducing the concept of schemas that govern motor processes. These schemas develop gradually throughout the aforementioned phases of motor learning. 
The two schema refer to the relationships, that users learn between the actual outcome of a movement and its response specification (recall schema) --- i.e., how a movement is executed --- and sensory consequences (recognition schema) --- i.e., how it felt~\cite{zwickerReflectionMotorLearning2009, schmidt19762schema}. Both schemas work in tandem and govern known movements as well as facilitate the learning of novel movements. Variability and repetition are key to develop strong schema responses~\cite{schmidt19762schema}.

\citet{schmidtMotorControlLearning2018}'s theory posits that learners form \GMPs{} rather than learning specific movements by altering parameters such as force or speed. 
Hence, allowing for variability and self-governed improvements develops robust GMPs.
Effective teaching therefore combines movement recall with opportunities for error recognition, supporting both schemas.

\subsection{Conceptualizing Feedback}
\label{subsubsec:KRKP}
Feedback can be categorized into \textit{knowledge of results (KR)} and \textit{knowledge of performance (KP)}~\cite{schmidtMotorControlLearning2018, salmoni1984knowledgeresults}. In essence, KR is goal-oriented and provides information about the outcome of a movement ("You missed the goal"). In contrast, KP is feedback about the movement pattern itself ("Your elbow was not straight"), a feedback form often employed by trainers, e.g., when correcting posture. Yet, both feedback forms usually coexist\footnote{For example, a trainer correcting throw form, yet also providing feedback on hit targets.}, and are --- at times --- even hard to discern as correct movement becomes the goal.

Regardless of its form, feedback is essential to accelerate motor skill consolidation~\cite{schmidtMotorControlLearning2018} allowing users to more accurately measure the actual outcome of the movement. The communicated error, presented as deviation from the desired state, consequently updates their motor response schema to form a generalized motor program more quickly.




\subsection{Attentional Focus Impacts Motor Learning}
\label{subsubsec:externalfocus}
Ultimately, providing learners with the right amount and type of information is crucial for skill acquisition. The method of delivering instructions and the focus of attention significantly affect learning effectiveness. \citet{wulfOptimizingPerformanceIntrinsic2016} introduced the OPTIMAL theory to address how motivational and attentional focus, either internal or external, impact motor learning. An internal focus relates to one's body movements, while an external focus emphasizes the intended effect of the movement~\cite{wulf2010instructionsmotor}. Research shows that feedback, particularly with an external focus, enhances learning and movement effectiveness more than an internal focus does~\cite{schmidtMotorControlLearning2018,sheaEnhancingMotorLearning1999}. 

\subsection{Improving Subskills Contributes to Overall Skill}
\label{subsubsec:subskill}
Insights into motor learning processes can be applied to enhance feedback systems, especially when designing for complex tasks, such as dancing. Motor learning theory recommends to divide a complex task into smaller chunks and teach them individually~\cite{krasnowMotorLearningControl2015, lindsayAdaptableCoachCritical2024}. 
For example, it can significantly help novice dancers to focus on one aspect at a time, combining them and adding more complex tasks later~\cite{krasnowMotorLearningControl2015, yangAutomaticDanceLesson2012}. Supporting them in just one aspect of dancing first will increase their overall proficiency in the long run. Additionally, learners can transfer subskills or parts of a complex task to other, similar ones after practicing and internalizing one first~\cite{schmidtMotorControlLearning2018}.

\subsection{Summary}
There are lots of theories about how people learn motor skills. To navigate them, we highlight specific concepts connected to dance movements here. Firstly, particular of interest for this work, is the schema theory~\cite{schmidtMotorControlLearning2018} and how users consolidate movements by establishing strong \GMPs{}.
As such, it is important for interactive dance support systems to facilitate the development of strong schema through repetition and appropriate feedback. Secondly, theory points us to appropriate feedback types for motor skill learning tasks. The type of feedback that is applicable also depends on the use case, such as providing \textit{Knowledge of Results} versus providing \textit{Knowledge of Performance}, or supporting the user with an external focus versus an internal focus of attention. We pursued the idea of using feedback to induce an external focus of attention as a promising approach for dancing.
Lastly, theories postulate that dividing complex task into smaller chunks is beneficial. As such, we focus our investigation on identifying subskills of dancing that are suitable to be represented in interactive dance systems.
\section{Practice: Exploring Opportunities for Interactive Systems to Support Dancing Practice}
\label{sec:e_interviews}
Dancing, fundamentally, is rooted in its social nature. Recognizing this, our design process aimed to explore the interplay of motor learning within a social setting. There is a correlation between self-controlled practice conditions and enhanced learners’ self-efficacy~\cite{schmidtMotorControlLearning2018}. Grounding our approach on this, we appreciated the motivational role feedback plays in the practicing process. With a vision of creating a system that fosters the ability to practice dancing, such as in one's home, we needed to understand the social context of dance. To make informed decisions, we initiated a series of interviews with five dance professionals. The primary goal of these interviews was to study the pedagogical strategies they employ and the tools that complement their teaching. These discussions not only provided us with an array of ideas for a potential system but also presented the challenges and limitations inherent to such designs. When identifying where an interactive system can be most beneficial, we sought to understand which specific subskills it can reinforce. Contrasting our approach with \citet{villaAssistingMotorSkill2021}, we concentrated particularly on the methods employed by instructors to impart skills of timing and rhythm.

\subsection{Procedure}
We conducted semi-structured interviews online via video conferencing software. In the beginning, the dance teachers provided informed consent for participation and recording. Then, the interviewer introduced the topic and the context of the interviews. The interviews comprised five phases: Demographics, Dance and Teaching Dance, Scenarios and Questions, Digital Support Tools: Experiences and Ideas, and Open Discussion. More details on the question can be found in \Cref{sec:interviewprotocol}. After collecting the demographics, we began the interview by querying dance teachers about their dance and teaching experience and their approaches to practicing dance. 
Next, we described scenarios showcasing different situations in class and how the teachers address student problems or difficulties. Subsequently, we discussed digital tools during teaching or dancing and what new tools they imagine to be helpful for practicing dance. Additionally, they elaborated on usable sensors and appropriate feedback to provide.

\subsection{Participating Dance Teachers}
\label{subsec:ei_participants}
We conducted a total of five interviews with different dance teachers (T1-T5). Three of the dance teachers were female, and two were male. The dance teachers were between 25 and 57 years old ($\samplemean=40.4$, $\samplesd=11.6$). Not all dance teachers were full-time dance teachers, but all had four or more years of experience ($\samplemean=19.0$, $\samplesd=10.7$). Together, they covered four different styles of dance. 
Each dance teachers was reimbursed with \$10 per hour.

\subsection{Findings}
\label{subsec:ei_results}
All interviews (6:16\,$h$) were recorded and transcribed verbatim. We followed the pragmatic approach to thematic analysis~\cite{blandfordQualitativeHCIResearch2016} to extract information from the interviews. A total of four researchers reviewed them and coded the statements of one representative interview. In an open discussion, we merged the codes, yielding the initial coding tree. Each reviewer then separately coded another interview based on that tree. Then, in an iterative discussion, we refined the coding scheme to build five themes that communicate key design aspects for systems that support dance practice.

Through the analysis, we identified the following recurring themes:
\begin{itemize}
    \item Components of Dance
    \item Teaching Methods
    \item Support Tools Used During Teaching
    \item Timing and Rhythm
    \item Problems or Risks for Technological Systems
\end{itemize}

\paragraph{Components of Dance}
The dance teachers agreed that aligning the movements to music, meaning the movement technique and the timing/rhythm, expressing something, bodily proprioception, and synchronization/harmony with the partner or group are the key components in dancing. In particular, they gave great importance to the component movement of music. It always consists of two parts (T3): movement, meaning having the correct execution and technique and timing these movements to the music, following its rhythm and beats. 
\paragraph{Teaching Methods}
In the second phase of the interviews, we asked the teachers about teaching methods for dancing to learn about a typical structure of dance classes. A regular dance class consists of a warm-up, teaching a new sequence, and a cool-down and stretching. Teachers first demonstrate a sequence, then break it into smaller parts, practicing them slowly and often without music. Movements for arms and legs are often taught separately. 
Counting beats (e.g., one to eight in 4/4 time) is commonly used to support rhythm learning. Even though one teacher does not like counting, she still does it as it helps the students.
\begin{quoting}
    If it’s a general problem that the person is basically not in-beat, then go through [the dance sequence] slowly without music, you can do it with counting. As I said, some people need the count from one to eight. (T3)
\end{quoting}
The teachers adapt the procedure to students' skill level and needs.
\paragraph{Support Tools Used During Teaching}
We queried the dance teachers about analog, mental, and technical support tools they used. Analog support tools include the mirror, gym benches, bars, scarves, and therabands, e.g. for partner coordination. 
Dance teachers further use metaphors and small stories to paint mental imagery of the respective movement and its transported emotion to support students in learning and expressing new movements. Technological tools primarily include music systems and streaming apps. 
Teacher 2 also used videos and slow motions as retrospective tools for performance analysis. They also utilized an application that determined a song’s speed and beats per minute.

\paragraph{Timing and Rhythm}
``Timing and Rhythm'' was a key concern in everyday dance pedagogy. To teach a feeling for the rhythm, teachers introduce it through simple movements, e.g., bouncing or letting students jump, walk, or stomp to  embody the beat.  
\begin{quoting}
    Coordination between clapping and stomping - stomping steps, is a successful method for me. (T5)
\end{quoting}
To align movement with music, teachers often use counting to structure start, duration and end of movements. 
Counting or clapping enhances the beat and makes it more prominant for the students (T2) and can be used with or without music.

As a reason for difficulties with timings, multiple teachers stated that mainly not a ‘lack of sense of rhythm’ caused timing problems, but the lack of motor skills and coordination:
\begin{quoting}
    For me, the rhythm thing is a discoordination. The lack of rhythm or the lack of dancing in-beat to the music of a person comes from a discoordination, from a disconnection between their head and body. (T5)
\end{quoting}
We further discussed suggestions for technological systems and feedback for beginner dancers, particularly the \textit{How?} and \textit{When?} to provide feedback. All teachers favored immediate, positive feedback, emphasizing that early correction prevents error consolidation and supports motivation, so T4.

Teachers suggested multiple feedback modalities ranging from visual to auditory and haptic feedback. One idea to practice the rhythm with visual feedback can be, for example, by using color changes (green/red) or pulsing lights indicating synchronization with the music.

\begin{quoting}
    There are actually mirrors where there is a light in the mirror or around the mirror, [...] and the mirror lights up red when you are no longer in sync, or for example, the mirror starts flashing in sync when you get out of sync. Then you know, when the mirror flashes, you have to concentrate specifically on the rhythm of the mirror again because you are no longer in rhythm with the song, so connecting it somehow. (T2)
\end{quoting}
Three of the dance instructors considered haptic feedback helpful, suggesting, for example, wearables such as a t-shirt or wristband that pulse with the beat of the music. 
Teacher 3 thought of a vibrating floor but noted it could be distracting, especially for elderly students.

Teacher 1 emphasized that haptic feedback is particularly helpful for problems regarding dance technique or motor skills, especially when directly applied to the respective body part to help locate errors. 
However, she considered vibrations on the body confusing. 

They also considered auditory feedback an option for rhythm training. Most teachers stated that once a detective system recognized that the dancer was out of sync, auditory feedback emphasizing the existing beat of the music should be introduced. This could be a metronome, or beat-counting and clapping as they practice it during their teaching. 
\begin{quoting}
    Assuming the mobile phone is also connected to the system, then [the system should add] an extra bass or beat to the existing beat. Because for some songs, it’s not easy to hear the beat; for example, you can get help at that moment, so a louder, additional beat sets in so one can orientate oneself again. (T2)
\end{quoting}
\paragraph{Problems or Risks for Technological Systems}
Lastly, we discussed the problems and risks of technological systems supporting dancers. All teachers mentioned that such systems cannot substitute dance lessons and teachers in real life and should rather be designed as a supplementary support tool. 
They also noted that unsupervised training may lead to injuries, for example when users skip essential warm-up and stretching routines. 
Using a support system could also lead to relying too much on its presence, highlighting the importance of providing diverse forms of support (T2).
\subsection{Implications and Connection to Motor Learning Theory}
\label{subsec:interview_implications}
Based on the insights from the interviews, we derived the following implications for interactive dance practice systems. We highlight how this established and empirically validated teaching methods are based on motor learning theory (see \Cref{subsec:theory}).

\paragraph{\textbf{Immediate Feedback to Consolidate Strong Schemas}}
Dance teachers try to provide immediate feedback whenever they observe errors to ensure no mistakes creep in, both on the students movements (cf. \textit{Knowledge of Performance}, \Cref{subsubsec:KRKP}) and correctness (cf. \textit{Knowledge of results}, \Cref{subsubsec:KRKP}) such as being in rhythm. This teaching approach directly focuses on developing strong schemas (cf. \textit{recall} and \textit{recognition} schema, \Cref{subsubsec:schematheory}) to establish dancing movements as a generalized motor program.

\paragraph{\textbf{External Focus of Attention Enhances Motor Learning}}
Teachers introduce an external focus of attention for the students through a well-established approach of clapping or counting the beat to support rhythm keeping. This method that aligns well with motor learning theory, showing that an external focus is more efficient (cf.~\Cref{subsubsec:externalfocus}).

\paragraph{\textbf{Tailoring Feedback to Motor Learning Phases}}
Teachers suggested on \textit{How?} and \textit{When?} to provide feedback for novice dancers. They adapt their teaching to the learners' needs. They provide more feedback in early learning stages (\textit{cognitive} and \textit{associative} stage, \Cref{subsubsec:phasesML}) and they reduce it when the students' dance proficiency increases (when transitioning to the \textit{autonomous} phase). Consequently, interactive feedback system should also dynamically adapt to the learner's progress.

\paragraph{\textbf{Rhythm as an Integral Subskill of Dancing}}
Teachers support the students in improving a subskill to master the overall skill (cf.~\Cref{subsubsec:subskill}) by splitting complex movements into smaller bits. They start by demonstrating the target movement completely once to give them a sense of the goal, and then show the parts of the movement slower and without music.


\section{\prototype{} - Supporting Rhythm Keeping Through Interactive Feedback}
\label{sec:designimpl}
After identifying design factors and requirements for rhythmic feedback for beginner dancers (see \Cref{sec:e_interviews}), we designed and implemented \prototype. \prototype{} consists of an instruction component using videos and a feedback component with clapping feedback. The components resemble real-world dance classes or video tutorials (instruction component) while adding auditory feedback on top (feedback component) to avoid clutter in the visual channel~\cite{sigristAugmentedVisualAuditory2013} and to follow proven teaching methods for rhythm feedback (cf. \Cref{subsec:ei_results}). The following section introduces the design of \prototype{} and its system architecture. 

\subsection{Design of \prototype}
\label{subsec:designPT}
In conceptualizing our interactive feedback system, \prototype{}, we identified five key design requirements (\textbf{D1-5}). These were derived from the teacher interviews and grounded in motor learning theory.

\begin{enumerate}
    \item[\textbf{D1}] \label{it:immFB} \textbf{Necessity of Immediate Feedback}: Drawing from the interviews with dance professionals (cf.~\Cref{subsec:interview_implications}), we recognized the need to support schema consolidation (cf.~\Cref{subsubsec:schematheory}) through \prototype. Our algorithm (see~\Cref{subsec:system}) is able to detect rhythm mistakes for each bar, offering sufficient granularity. Further, it is computationally lightweight, being able to decide on the necessity of feedback without delay.

    \item[\textbf{D2}] \label{it:ExtFoc} \textbf{Facilitate External Focus of Attention}: Dance teachers effectively use external focus of attention for rhythm-keeping feedback (cf.~\Cref{subsec:interview_implications}). The feedback implemented through \prototype{} directly follows these methods, mimicking a clapping teacher to help students to keep in rhythm. 

    \item[\textbf{D3}] \label{it:subskill} \textbf{Focus on One Aspect of Dance}: Teachers first focus on training individual parts separately, improving those subskills to further dancing proficiency as a whole (cf.~\Cref{subsubsec:subskill}). The interviews revealed that it is particularly important for dancers to understand when a move begins, how long it lasts and when it ends. According to the dance teachers, problems to get the timing right are caused by the lack of motor skills and coordination (cf. \textit{Timing and Rhythm}, \Cref{subsec:ei_results}). \prototype{} follows this concept of separating different aspects of dance and focuses on timing and rhythm as one component of dance practice. 
    The system introduces movements gradually and supports subsequent execution through rhythmic feedback.

    \item[\textbf{D4}] \label{it:timingFB} \textbf{Delivering Feedback at Appropriate Moments:} Different phases and stages of motor learning warrant different type and frequency of teacher support (cf.~\Cref{subsec:interview_implications}). With \prototype{}, students can practice at their own pace and repeat parts of the exercise as often as necessary. In addition, the rhythm feedback dynamically adjusts to the students' progress and only activates if the students needs it.

    
    
    \item[\textbf{D5}] \label{it:SoloPrac} \textbf{Self Practice Capability and Easy Deployment}: 
    To ease initial entry, \prototype{} relies on a simple camera-based rhythm detection method (see~\Cref{subsec:system}) allowing fast and easy deployment. The system only requires a webcam and speakers to record the dancer and provide rhythm feedback. 
\end{enumerate}


\subsection{Practicing with \prototype}
\label{subsec:practicing}
When practicing with \prototype{}, the user has to set up the webcam facing at the spot where the user will dance. Then the user only has to turn on the speakers and start the application. The system does not require calibration and is computationally lightweight to provide immediate feedback. The application guides the user through the individual steps of the exercise. The tutorials resemble dance tutorials on YouTube and incorporate the teaching methods we derived from the dance teacher interviews. We deliberately chose a video tutorial for the instruction of the dance movements and the clapping sound for the feedback to not add more clutter to the visual channel and to follow proven teaching methods for rhythm feedback~\cite{rahebDanceInteractiveLearning2019, sigristAugmentedVisualAuditory2013}.
One of the co-authors recorded the video following the insights from the dance teacher interviews. The video explains each dance move and step slowly and without music, putting the parts together eventually and adding the music at the very end. When dancing without music, the instructor in the video counts the beats to highlight when the step is intended to start and finish. For one training session the user has 15 minutes of practice time, including the video tutorial. This offers enough time to practice the moves and experience the interactive feedback. Whenever the user needs to revisit different parts, they can return to the video. After 15 minutes of practice time, the user can test their rhythm performance in a test session, where the feedback is not available. In the test the user performs the practiced dance sequence throughout the song.

\subsection{System Implementation}
\label{subsec:system}
\begin{figure}[h]
    \centering
    \includegraphics[width=.8\textwidth]{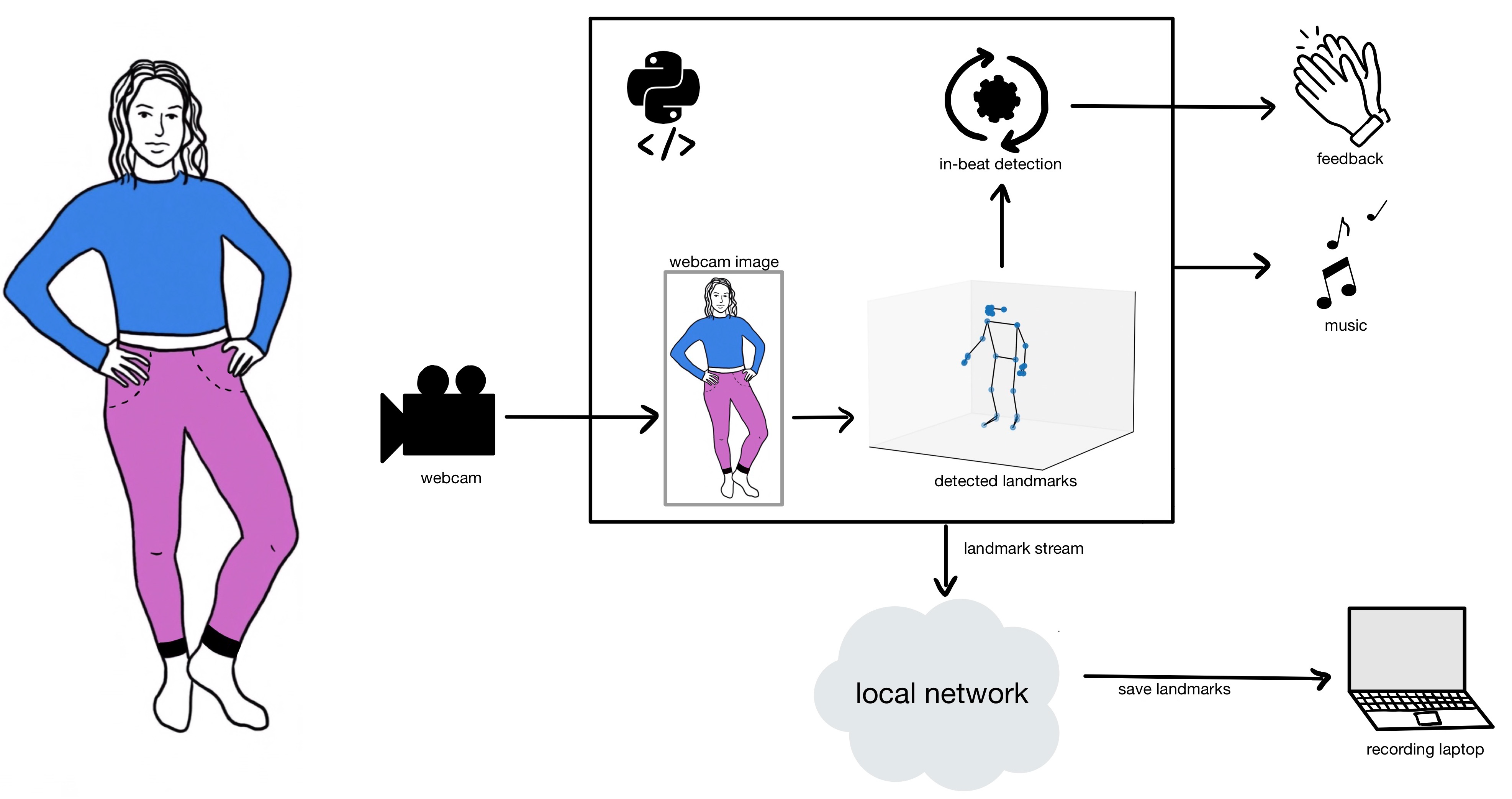}
    \caption{Software architecture of \prototype{}, detailing the detection algorithm.}
    \label{fig:architecture}
    \Description{Architecture of SkeletonDance. A person is recorded by a webcam. The webcam images are analyzed by a Python algorithm. The algorithm detects landmark data by extracting the skeleton from the image. The landmark data is streamed into a local network and saved on the recording laptop. The detected landmarks are used for the continuous in-beat detection, which then results either in clapping feedback in addition to the music when the person is not in sync with the beat or the music keeps playing normally.}
\end{figure}

\prototype{} consists of several modules that are interconnected through a local network: (1) a webcam-based algorithm to detect the user's current pose, (2) a subsequent algorithm that assesses the necessity of feedback, and (3) a graphical user interface guiding users through the training program. A complete overview is provided in \Cref{fig:architecture}. The system is directly tailored to \textbf{D5}, providing a system that is easy to deploy and allows for solo practice. For an overview of other systems and recording alternatives, see~\Cref{subsec:ids_fb}.

\subsubsection{User Pose Detection}
\label{subsec:pt_signal_acqu}
We used an off-the-shelf webcam (see~\Cref{subsec:s_setup}) to record body movements. Using OpenCV\footnote{\url{https://opencv.org/}} and the Mediapipe pose-estimator\footnote{\url{https://developers.google.com/mediapipe/solutions/vision/pose_landmarker/}}, we extracted a skeleton for the dancer at 20 frames per second. The skeleton includes thirty-three landmarks, including major human body joints, and facial features. Recording the data over time allowed us to track the dancer without compromising privacy. No images from the webcam were saved or used for further processing. The skeleton data was streamed into the local network using lab streaming layer\footnote{\url{https://github.com/sccn/labstreaminglayer}}, ensuring time synchronization.

\subsubsection{Algorithm for In-Beat Detection}
\label{subsec:pt_algo}
We implemented a lightweight and robust detection of the employed dance sequences (see~\Cref{subsec:studydesign}). \prototype{} is aimed at beginner dancers; thus, an elaborate detection of a multitude of dance movements was not necessary. Consequently, we intentionally opted for a lightweight rather than a sophisticated algorithmic.

\begin{figure}[htb]
    \begin{minipage}[b]{0.45\linewidth}
            \centering
    \includegraphics[width=.4\textwidth]{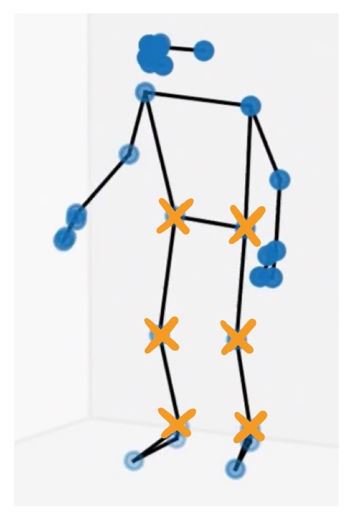}
    \caption{Landmarks of the skeleton used for In-Beat Detection. }
    \label{fig:landmarks_inbeat_detection}
    \Description{Skeleton visualization of detected landmarks of the body. For the in-beat detection, only the landmarks of the hips, knees, and ankles are used.}
    \end{minipage}
    \hspace{0.5cm}
    \begin{minipage}[b]{0.45\linewidth}
            \centering
    \includegraphics[width=\textwidth]{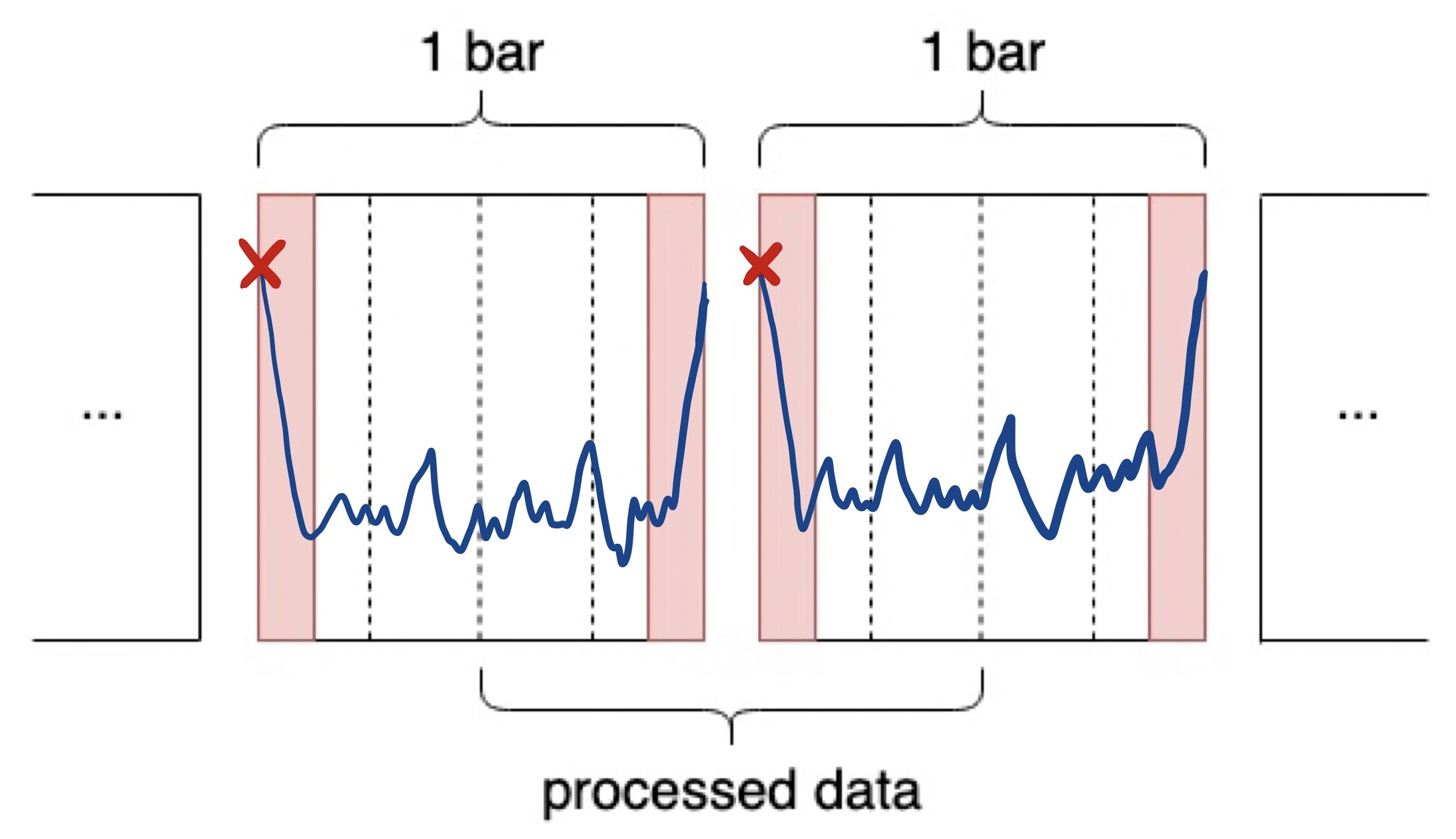}
    \caption{Bar sequence and associated \distancemetric{} over time. The red cross marks the detected peak (if the dancer is in rhythm). Note that the data processing itself is shifted by half a bar.}
    \label{fig:sliding_window_detection}
    \Description{Graph depicting leg distance over time separated into two windows of length one bar each. At the start and end of each bar, the leg distance is high and visually highlighted in red.}
    \end{minipage}
\end{figure}

Using the distinct movements of lifting up the legs at beat 1 of a song's bar (see~\Cref{subsec:studydesign}), we can define a distance metric for these landmarks. More specifically, we calculated the Euclidean distance between the hip joint, the knee, and the ankle for each leg (see \Cref{fig:landmarks_inbeat_detection}) and summed them up in a total distance metric \distancemetric{} per frame.

Our detection algorithm runs for every bar of the song and assesses if the dancer is out of rhythm. Given a song with 99 beats per minute (bpm)\footnote{Song used in this study ``Summer Nights (Tropical House Music)'' \url{https://www.youtube.com/watch?v=xGtIWbY9dTU} adjusted to 99 bpm.}, one bar approximately lasts 2.42 seconds. Using a 1-bar wide window, we extract 48 frames for which we calculate \distancemetric{} each. If the dancer is in rhythm, we expect a large distance value at the start (or end respectively) of each bar. Running a peak detection on individual bars allowed us to determine whether the highest peak (corresponding to lifting the leg) was synchronized with the song's rhythm. In this case, it can only be found at the start (or end, respectively) of a bar. Note that the data processing itself was shifted by half a bar to elicit one peak response per window (see \Cref{fig:sliding_window_detection}).

\subsubsection{Technical Feasibility Assessment}
\label{subsec:system_evaluation}
To assess the technical feasibility of the detection algorithm, we conducted preliminary testing prior to the study. Three users performed the target dance sequences while deliberately mimicking on-beat, early, late, and off-beat movements. These trials were systematically observed, and the experimenter (an experienced dancer) verified that the system’s detections aligned with the intended timing deviations. Importantly, our goal was not to achieve perfect or exhaustive detection accuracy. As \prototype{} is designed as a supportive system rather than a strict assessment tool, occasional inaccuracies are justifiable and do not undermine its usefulness.
To further validate correct system behavior under realistic conditions, we conducted two pilot studies following the same procedure described in \Cref{subsec:s_procedure}, with the experimenter present to confirm that the system worked as intended.

\subsubsection{Integrating Feedback for the Dancer}
\label{subsec:pt_feedback}
After detecting rhythm irregularities, we introduce feedback to support the dancer. This feedback is based on the implications derived from the motor learning theory and the interviews (cf.~\Cref{subsec:interview_implications}).
It follows traditional teaching methods by incorporating a background clapping on every beat. A voting algorithm ensured a smooth experience for the dancers by only activating the clapping feedback when the dancer was out of rhythm for three consecutive bars. Deactivating the feedback happened in an analogous way.

\begin{figure}[h]
    \centering
    \includegraphics[width=.7\textwidth]{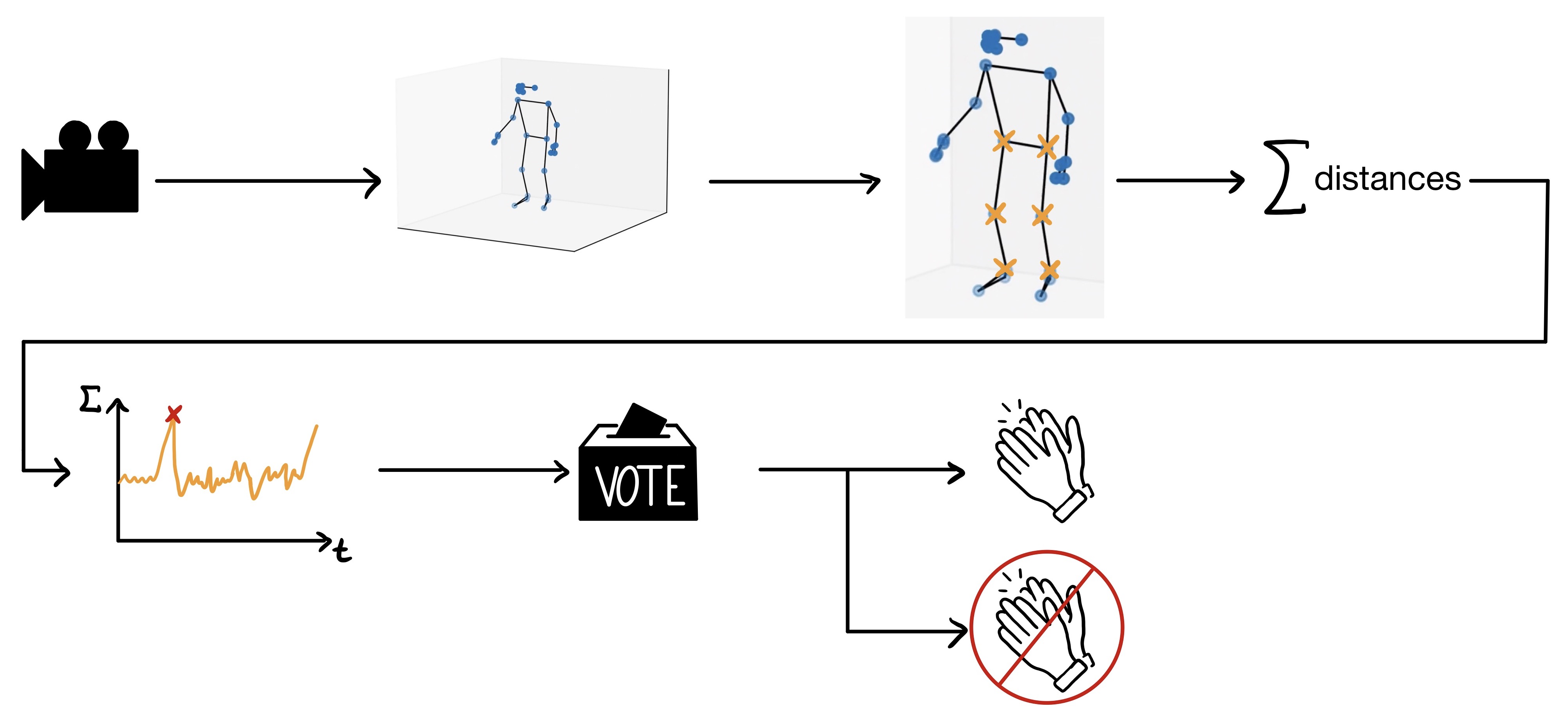}
    \caption{Overview of \prototype's processing pipeline, showcasing how the system arrives at the final decision to activate (or deactivate) feedback for the dancer. }
    \label{fig:algorithm}
    \Description{Overview of SkeletonDance’s processing pipeline, showcasing how the system arrives at the final decision to activate. The webcam records images, OpenCV and the Mediapipe pose-estimator extract the landmarks of the skeleton. For further analysis, we only use the landmarks of the hips, knees, and ankles. The algorithm calculates the Euclidian distance for these landmarks for each leg and sums them up to a total distance metric per frame. For a one-bar window, we expect a large distance value at the start (or end respectively) of a bar, displayed as a high peak in the data. A voting algorithm determines whether the person is in sync with the beat of the music and activates or deactivates the clapping.}    
\end{figure}

An overview of the full processing pipeline, including pose estimation, \distancemetric{} calculation, and feedback classification, is shown in \Cref{fig:algorithm}.

\section{Evaluation} 
\label{sec:study}

To gain insights into the acceptance, usability, and constraints of \prototype{}, we evaluated \prototype{} in a user study. The participants practiced two dance sequences and trained them without feedback or experienced the feedback from \prototype{}.

\subsection{Study Design}
\label{subsec:studydesign}
\begin{figure}[h] 
    \centering
    \includegraphics[width=\textwidth]{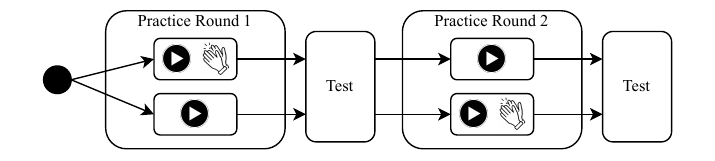}
    \caption{Study procedure. The arrows show a path through the study. For example, if the participant started with ``video and clapping feedback'' in the first round, they received ``video only'' in the second round. During the test session, the participants had no video instruction and no feedback.}
    \label{fig:timeline}
    \Description{Flow graph of the study procedure with the different conditions per round. The Figure shows a starting point followed by a box for the first practice round with the two different feedback types “video + clapping” or “video only” feedback.  After the first practice round, the participants conduct a test without any feedback. After that, they start the second practice round with swapped feedback and a second test afterwards. }
\end{figure}

We employed a within-subject study design with the video instruction and the clapping overlay in the feedback condition (\fb) or only the video instruction as the baseline condition (\base). 
The participants were asked to practice two simple dance sequences, one for each condition. Each dance sequence was composed of three different but similar dance moves. One dance move included a sequence of four steps throughout one 4/4 bar using the distinct movement of lifting the legs at beat 1 of a bar for the detection algorithm. The dance moves are independent of a specific dance style. We chose six similar moves to ensure that the two sequences are comparable. All of the moves were leg movements and easy to practice. We deliberately chose not to include the arms, making the dance more suitable for beginner dancers by reducing the complexity to one component, as suggested in motor learning theories~\cite{krasnowMotorLearningControl2015}. The dance sequences were chosen in consultation with one of the experts to be of the same difficulty. We chose the song and the dance moves independent of a specific dance style but they are exemplary for dances with regular rhythms~\cite{karageorghisEffectsAuditoryRhythm2019}.
We counterbalanced the order of the conditions and the dance sequences the participants had to practice. After training the sequence, we tested the participants so they performed the practiced sequence throughout one playback of the song without feedback. The procedure with the different feedback conditions is depicted in~\Cref{fig:timeline}. We calculated rhythmic accuracy metrics from the landmark data to quantitatively assess the participants' dance performance as detailed in \Cref{subsec:metrics}. We further used established and customized questionnaires listed in \Cref{tab:custom_questions} to get insights into the participants’ opinions towards \prototype{} after each practice round. To gain further insights, we conducted semi-structured interviews, allowing them to report more details on their experiences with the system. 

\subsection{Apparatus}
\label{subsec:s_setup}
\begin{figure}[htb]
	\centering
	\includegraphics[width=.4\textwidth]{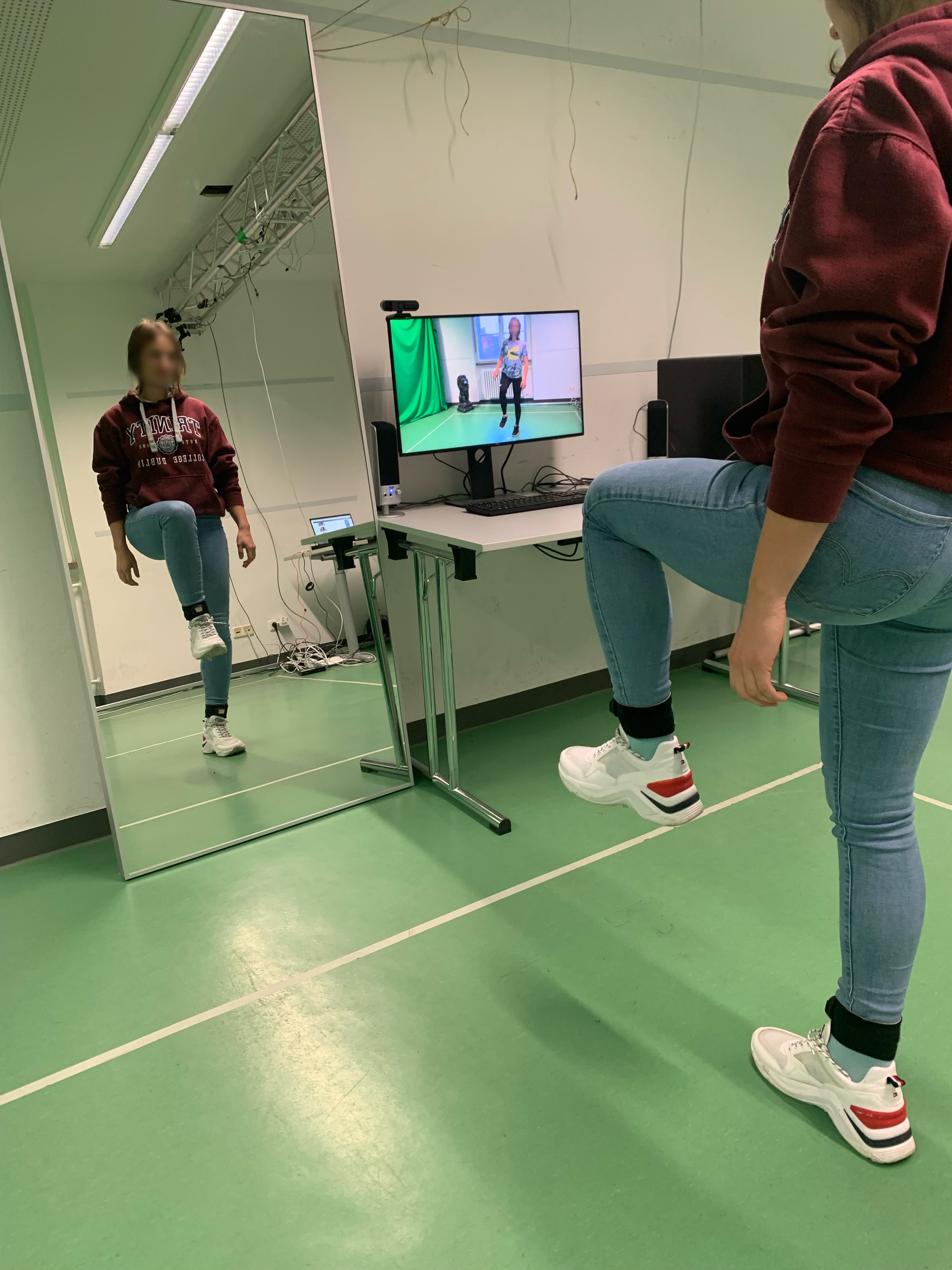}
	\caption{Participant in the study setup, following along the video tutorial while receiving visual feedback from the mirror. The music is playing, and the webcam is recording. }
 \label{fig:study_apparatus}
 \Description{A photo of a user learning a new dance sequence. The woman stands in front of a mirror and lifts her left leg. Next to the right of the mirror is a table with a monitor, a webcam on top, and two speakers. The monitor displays the video tutorial with the instructions. }
\end{figure}

\Cref{fig:study_apparatus} shows the setup during the study. We included a mirror, which is common in dance practice, to provide visual feedback so that the participants can observe themselves. We placed a computer screen next to the mirror to display the \prototype{} application in full-screen mode. 
We placed some speakers for the audio playback next to it. We put the camera on top of the screen close to the mirror for a frontal recording. 
The camera recorded the participants during the dance sessions. The program turns the images into stick figures. Only the stick figure is being recorded and used for analyzing the body movements. We do not store images or videos. 
The recording PC is not displayed in \Cref{fig:study_apparatus} as it was placed behind the participant on an additional desk and is only used by the experimenter to run the program. For technical details about the system and the algorithm, please refer to~\Cref{subsec:system}.
\subsection{Procedure}
\label{subsec:s_procedure}
In the beginning, the experimenter welcomed the participant and described the study procedure. 
After they provided informed consent, we used a questionnaire to collect demographics and information about their dance and rhythm experience. Additionally, we tested the participant's rhythmic abilities in a short rhythm test. For the test, we used a 10-second excerpt from the song they later danced to and asked them to clap to every beat of the music. Later, three experts checked the rhythm test recordings to identify whether participants showed serious problems syncing their clapping to the beats. 
Clapping sounds are common musical features and could be mistaken as part of the actual music. To prevent that, we also showed them an audio sample demonstrating what the music with the auditory feedback sounds like.

After the rhythm test, we informed the participants about the following steps in the study, which consisted of two times practicing a dance sequence in either the \base{} or \fb{} condition. 
After ensuring the functionality of the system and the recording PC, the experimenter left the room to simulate a more realistic home dance training and to prevent participants from being distracted or intimidated.

The participant consequently started practicing the first dance sequence under the particular condition following the video tutorial. The video tutorial explained the dance moves slowly, first without music, and added the music at the end (cf.~\Cref{subsec:practicing}). 
The participants had 15 minutes to practice with the video tutorial and repeat it as often as they wanted during this time. One video lasted 7:43 minutes, and the other one 7:53 minutes. Once the 15-minute timer was up, they conducted the test without the feedback independent of the condition by dancing the practiced sequence throughout one playback of the song. 
Upon completion, the experimenter returned to the room and gave a set of questionnaires to the participant to complete on a tablet. We used a set of custom questions (\Cref{tab:custom_questions}) and standardized questionnaires to gain insights about the workload using NASA TLX~\cite{hart1988NASATLX}, Flow Short Scale~\cite{rheinbergErfassungFlowErlebens2003}, self-appraised performance rating~\cite{mamassis2004effectsmental}, and creepiness scale~\cite{woźniakCreepyTechnologyWhat2021}. After finishing the questionnaires, the experimenter left the room again, and the participants started the second practice round according to the same scheme as in the first round but with the other conditions and dance sequence. When they finished, the experimenter asked them to answer the questionnaires again. The second questionnaire had two additional questions targeting the participant’s prior knowledge of the dance sequences and their perceived difficulty. We then conducted a semi-structured interview to inquire about the participants' experience during the study. he interview contained questions about their general impression, whether the system served their needs, whether the feedback was noticeable, and whether it provided new insights, potential user groups, improvements, and suggestions. In total, the study lasted approximately $90\,min$.

\begin{table}[htb]
	\caption{Custom questions after each control modality, targeting perceived timings and perception of rhythm. All rated on a visual analog scale (VAS) from 0 to 100; strongly disagree to strongly agree.}
	\label{tab:custom_questions}
	\centering
	\begin{tabular}{ll}
			\toprule
			\multicolumn{2}{l}{\textbf{Custom questions regarding the system and its feedback.}}\\
			\midrule
			\textbf{Q1} & I noticed the system while dancing.\\
			\textbf{Q2} & The system was helpful to me.\\
			\textbf{Q3} & The system helped me to gain new insights into my timing while dancing.\\
			\textbf{Q4} & The system allowed me to reflect on my timing in regards to the music.\\
			\textbf{Q5} & I used the system to adapt to the rhythm.\\
			\textbf{Q6} & I used the system to find back into the rhythm.\\
			\textbf{Q7} & The system was in line with my body perception.\\
			\textbf{Q8} & The system was shown to me long enough to process it.\\
			\bottomrule
		\end{tabular}
\end{table}

\subsection{Metrics and Analysis}
\label{subsec:metrics}
We evaluated \prototype{} using a combination of objective motion data from the camera recordings and subjective data in the form of questionnaires and semi-structured interviews.
\subsubsection{Objective Metrics of Dance Performance}
\label{subsec:objective_metrics}
Based on the landmark data from the camera recordings collected throughout the study, we calculated a series of metrics to assess the participants' rhythmic accuracy~\cite{diaspereiradossantosYouAreBeat2018} and dance performance. All metrics are calculated solely based on the test session after practicing a dance sequence (either with \fb{} or \base).
\begin{itemize}
    \item \emph{\Beatshit} measured as ratio of in-rhythm bars to total bars.
    \item \emph{\Amtlost}: amount of times a participants lost the rhythm. Only counted occurrences, i.e., subsequent out-of-rhythm bars are not penalized. Actively counted once a participant started dancing and only considered out-of-rhythm occurrences if rhythm was regained afterwards, making the metrics robust against participants who began later and disengaged prematurely.
    \item \emph{\Meanofftimes}: Mean duration (in bars) it took participants to regain the rhythm. Analog to \emph{\amtlost}, this metric was protected against participants who began later and disengaged prematurely.
    \item \emph{\Fbduration}: Duration (in seconds) of clapping feedback that participants experienced \textit{during their training session}. Note that this metric was only available for exactly one training session, in which participants were assigned the \fb{} condition. This is a supporting metric that does not measure participants' dance performances but rather assesses the intended functionality of \prototype, cf.~\Cref{subsec:system_evaluation}.
\end{itemize}

For all metrics, we opted to include the participants' dancing experience as a moderator variable when conducting our analysis. We fitted individual linear mixed-effects models predicting each metric and reported significant effects. As fixed effects, we used condition, dancing experience, and a by-participant random effect. For each model, we confirmed that the necessary assumptions of the model were met, indicating no substantial violations of homoscedasticity, linearity, normality of residuals, or independence. In case of multiple comparisons, we report Tukey-adjusted p-values.

\subsubsection{Subjective Metrics}
We used a set of custom questions (\Cref{tab:custom_questions}) inquiring about whether \prototype{} supported users in their rhythmic perception, in particular if it allowed participants to find back into the rhythm, aligned with their own bodily perception and whether the system was noticeable. Further standardized questionnaires queried workload using NASA TLX~\cite{hart1988NASATLX}, flow experience through the Flow Short Scale~\cite{rheinbergErfassungFlowErlebens2003}, self-appraised performance rating~\cite{mamassis2004effectsmental}, and perceived creepiness of \prototype{} through the creepiness scale~\cite{woźniakCreepyTechnologyWhat2021}. Analog to our objective metrics, we analyzed all questionnaire using linear-mixed effects models. See \Cref{subsec:objective_metrics} for details.

Concluding semi-structured interviews allowed the participants to report more details on their experiences with the system. The interview contained questions about their general impression, whether the system served their needs, whether the feedback was noticeable, and whether it provided new insights, potential user groups, improvements, and suggestions. We conducted the interviews either in English or German according to the participant's preference. They were all recorded and transcribed verbatim. For the analysis, we followed the same approach to thematic analysis~\cite{blandfordQualitativeHCIResearch2016} as detailed in \Cref{subsec:ei_results}.

\subsection{Participants}
\label{subsec:s_participants}
We recruited $26$ participants through university mailing lists and word of mouth. We specifically recruited participants who have little or no dancing experience or say about themselves that they are not good dancers. For the study participation, we instructed them to wear comfortable clothes that allow them to lift up the leg at a 90-degree angle between the upper and lower leg. We excluded two participants due them not completing the study procedure and used the data of $24$ ($14\,m, 9\,f, 1\,divers$; age $\samplemean=25.58\,y, \samplesd=2.65\,y$) for the analysis.
Regarding their previous dancing experience, participants reported 54\footnote{\label{footnote:vas_experience}Visual analog scale (VAS) from 0 to 100, "I have no experience" to "I have a lot of experience."} as the highest rating ($\samplemean=24.42$, $\samplesd=16.77$). They rated their rhythm experience higher\footref{footnote:vas_experience} ($\samplemean=43.88$, $\samplesd=28.10$). Dance and rhythm-related activities reported by the participants ranged from none, over club dancing, juggling, cycling, and cardio with rhythm, to Zumba, playing different instruments (guitar, ukulele, piano), and rap. 
The dance sequence had a medium high difficulty ($\samplemean=62.3$, $\samplesd=20.7$) for the participants and they were not familiar with it beforehand ($\samplemean=25.2$, $\samplesd=32.3$).
Each participant was reimbursed with \$10 per hour. Ethical approval for the study was obtained from the review board of the first author’s affiliation.

\subsection{Results}
\label{sec:s_results}
In the following, we report on the statistical analysis of our collected metrics, containing the analysis of the objective metrics, questionnaires and interview answers. Please refer to \Cref{subsec:metrics} for details on the analysis method. 
\subsubsection{Objective Metrics}
\label{subsec:s_obj_results}
We collected landmark data during the test session of the study. Participants did not receive feedback during this phase. The distinction of \base{} and \fb{} refers to the condition they experienced in the practice round before the corresponding test session.

\begin{figure}
    \centering
    \includegraphics[width=.8\textwidth]{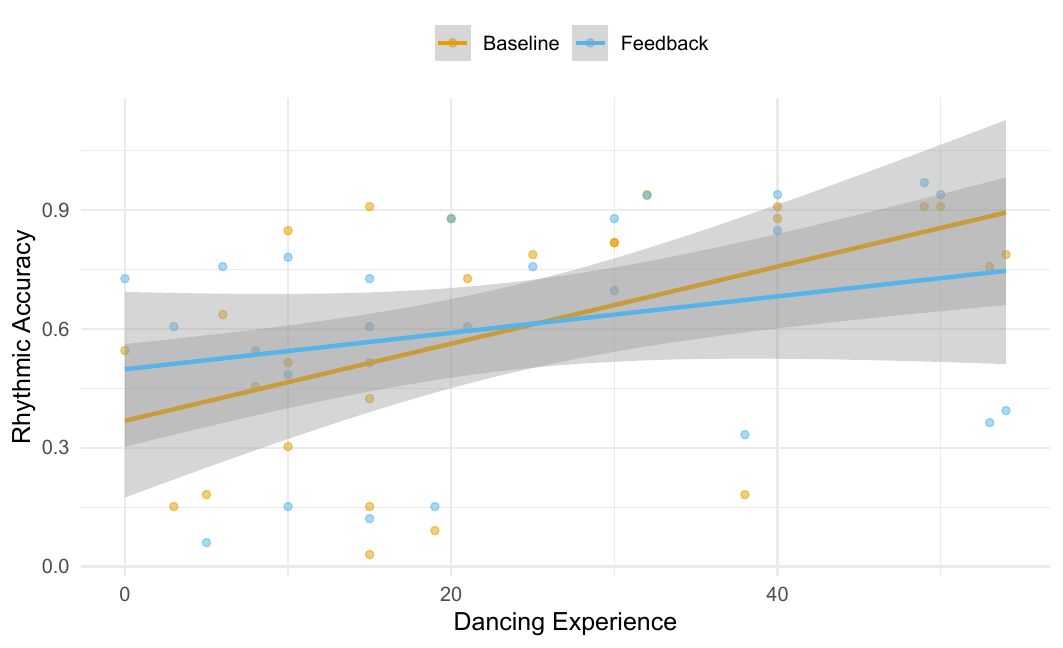} 
    \caption{\Beatshit{} in relation to dancing experience as predicted by our model. Feedback leads to a higher \beatshit{} for participants with low dancing experience. However, the \fb{} condition exhibits a weaker positive slope given increasing dancing experience than \base, indicating a cross-over point beyond which feedback becomes less effective.}
    \label{fig:beatshit_danceexp}
    \Description{Rhythmic Accuracy in relation to dancing experience as predicted by our model. Feedback leads to a higher \beatshit{} for participants with low dancing experience. However, the \fb{} condition exhibits a weaker positive slope given increasing dancing experience than \base, indicating a cross-over point beyond which feedback becomes less effective.}    
\end{figure}

\paragraph{\Beatshit}
The grand mean of \beatshit{} was $60.2\%$ ($\samplesd=29.2\%$). Participants in the \base{} condition had a lower \beatshit{} ($\samplemean=59.8\%$, $\samplesd=31.0\%$) than participants in the \fb{} condition ($\samplemean=60.7\%$, $\samplesd=27.8\%$). 
We found a significant main effect of dancing experience ($\beta = 0.010$, $\stderror = 0.003$, $t(28.2) = 3.01$, $p = .006$), indicating that higher dancing experience was associated with higher \beatshit. This effect was moderated by condition as shown by a significant interaction effect between condition and dancing experience ($\beta = -0.005$, $\stderror = 0.002$, $t(22) = -2.09$, $p = .048$). No significant main effect of condition was found. 
Simple-slope analysis showed a positive association between dancing experience and \beatshit{} in \base{} ($\beta = 0.010$, $\stderror = 0.003$, $t(28.2) = 3.01$, $p = .006$), whereas the slope was smaller and not significant in \fb{}. The difference between slopes was significant ($\beta = 0.005$, $\stderror = 0.002$, $t(22) = 2.09$, $p = .048$), indicating an attenuation of the interaction in the Feedback condition. The interaction is depicted in \Cref{fig:beatshit_danceexp}, which shows marginal effect plots of predicted \beatshit{} as a function of dancing experience for both conditions. 
In line with the statistics, we observed a higher \beatshit{} with increasing dancing experience. Participants with a lower dancing experience had a higher \beatshit{} in the \fb{} condition compared to \base. We can also observe the differing slopes resulting in a cross-over point. Beyond this point \fb{} results in a lower \beatshit{} than the \base{}. 

\begin{figure}
    \centering
    \includegraphics[width=.8\textwidth]{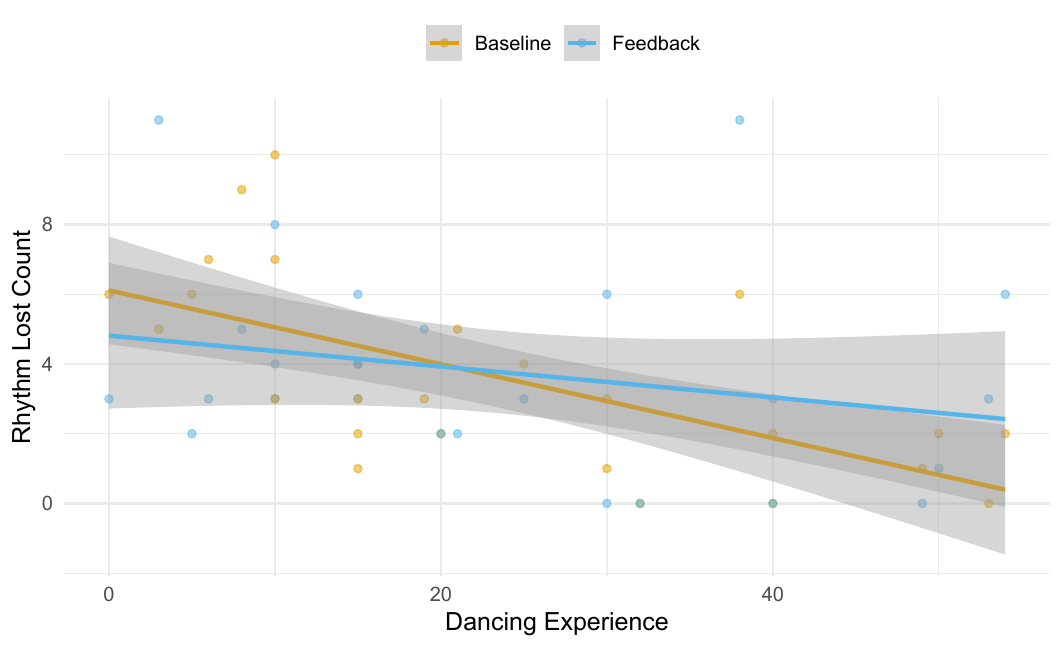} 
    \caption{\Amtlost{} in relation to their dancing experience as predicted by our model. Participants with low dancing experience lost the rhythm more frequently in the \base{} condition than in the \fb{} condition. With increasing dancing experience rhythm loss decreased more strongly in the \base{} condition than in \fb{}, resulting in a cross-over point after which feedback becomes less effective.}
    \label{fig:amountlost_danceexp}
    \Description{Rhythm Lost Count in relation to their dancing experience as predicted by our model. Participants with low dancing experience lost the rhythm more frequently in the \base{} condition than in the \fb{} condition. With increasing dancing experience rhythm loss decreased more strongly in the \base{} condition than in \fb{}, resulting in a cross-over point after which feedback becomes less effective.}
\end{figure}

\paragraph{\Amtlost}
For \amtlost, the grand mean was $3.69$ ($\samplesd=2.82$). Participants in the \base{} condition lost it on average $=3.62$ times ($\samplesd=2.74$), participants in the \fb{} condition $=3.77$ times ($\samplesd=2.96$).
We found a significant main effect of dancing experience ($\beta = -0.101$, $\stderror = 0.033$, $t(38.02) = -3.08$, $p = .004$), indicating that higher dancing experience leads to fewer rhythm losses. No further significant effects were found. Albeit not significant, we provide an interaction plot of dancing experience and condition in \Cref{fig:amountlost_danceexp}. Consistent with our expectations, this metric shows a pattern inverse to \beatshit. Participants with little dancing experience had a higher rate of rhythm loss, while higher dancing experience is associated with a lower loss rate overall. For participants with low dancing experience, rhythm was lost more frequently in the \base{} condition than in the \fb{} condition. However, this relationship reverses with increasing dancing experience. For participants with higher dancing experience, rhythm was lost less often in the \base{} condition compared to \fb.

\begin{figure}
    \centering
    \includegraphics[width=.8\textwidth]{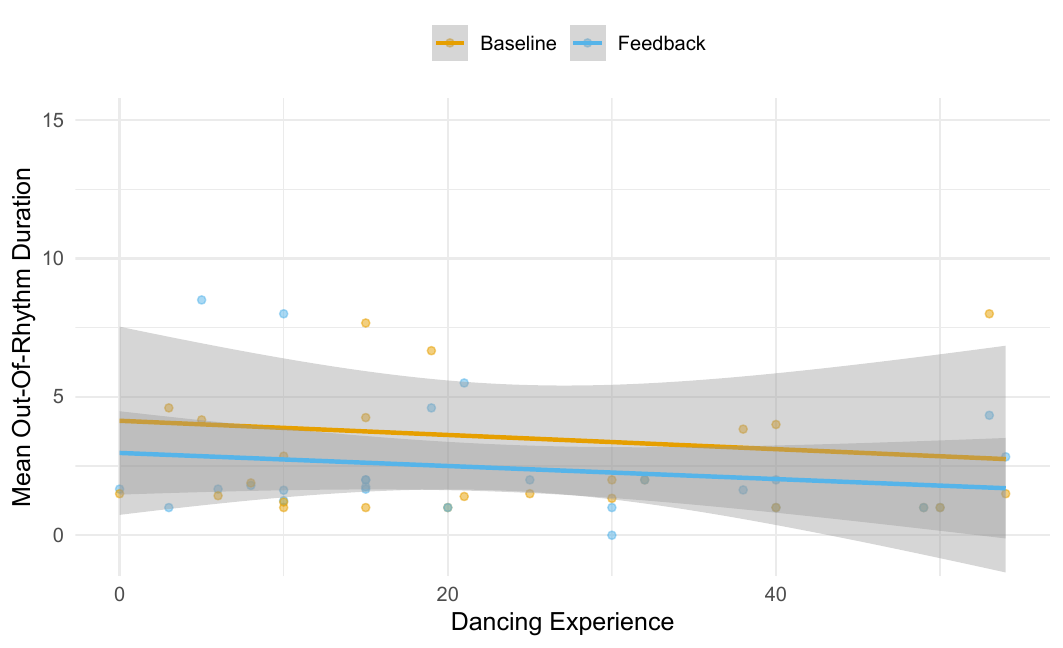}
    \caption{\Meanofftimes{} (in bars) are longer in the \base{} condition than in the \fb{} condition across all levels of dancing experience. Overall, increasing dancing experience is associated with slightly shorter \meanofftimes{}, indicating faster rhythm recovery.} 
    \label{fig:meanoff_danceexp}
    \Description{Mean Out-Of-Rhythm Duration (in bars) are longer in the \base{} condition than in the \fb{} condition across all levels of dancing experience. Overall, increasing dancing experience is associated with slightly shorter \meanofftimes{}, indicating faster rhythm recovery.}    
\end{figure}

\paragraph{\Meanofftimes}
The grand mean of \meanofftimes{} was $2.97$ bars ($\samplesd=3.63$). Participants in the \base{} condition lost the rhythm for more bars ($\samplemean=3.53$, $\samplesd=4.67$) than participants in the \fb{} condition ($\samplemean=2.42$, $\samplesd=2.10$). Our statistical analysis found that neither the main effects of condition or dancing experience, nor their interaction, were significant.
Overall, \meanofftimes{} is higher for the \base{} condition than for the \fb{} condition, independent of the dancing experience (see \Cref{fig:meanoff_danceexp}). For both conditions, we observed a slight decrease of \meanofftimes{} with increasing dancing experience, indicating that more experienced participants recovered the rhythm more quickly. Shorter \meanofftimes{} therefore indicate a quicker recovery once the rhythm was lost.

\begin{figure}
    \centering
    \includegraphics[width=.8\textwidth]{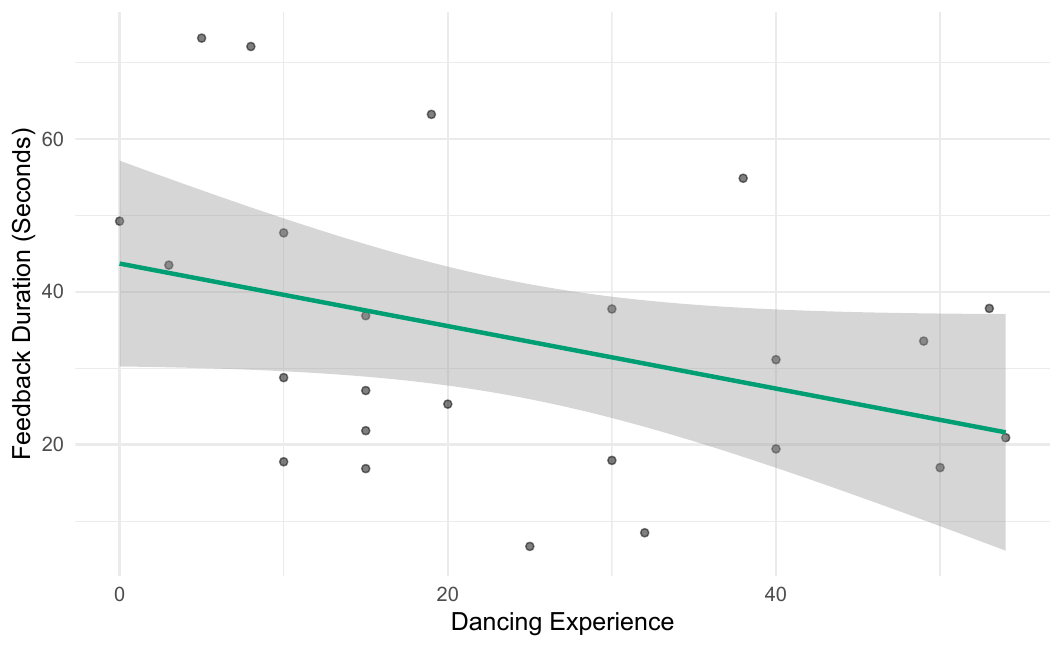}
    \caption{\Fbduration{} in relation to the dancing experience in the \fb{} condition during the practice session. For participants with lower dancing experience the feedback was active for a longer duration. With increasing experience, the feedback duration decreased.} 
    \label{fig:fbduration_danceexp}
    \Description{Feedback Duration in relation to the dancing experience in the \fb{} condition during the practice session. For participants with lower dancing experience the feedback was active for a longer duration. With increasing experience, the feedback duration decreased.}    
\end{figure}

\paragraph{\Fbduration}
Additionally to the reported metrics during the test session, we also measured (during the practice phase in the \fb{} condition) for how long the clapping feedback was active (in seconds). This supporting metric provides insight into the correct and intended functionality of \prototype. In other words, whether the system supported those likely to struggle more (low dancing experience) more often than participants with a higher dancing experience.
The feedback was on average active for $\samplemean=33.5$ seconds ($\samplesd=24.4$). Statistical analysis (without condition as a fixed effect) revealed no statistical significant effects. The relationship between the feedback duration and the dancing experience is depicted in \Cref{fig:fbduration_danceexp}. As anticipated, for participants with lower dancing experience the feedback was active longer, whereas increasing dancing experience was associated with a shorter feedback duration, indicating that more experienced participants relied less on clapping feedback, being able to stay in rhythm more accurately.

\subsubsection{Questionnaires}
\label{subsec:s_quest_results}
\begin{figure}[htb] 
    \centering
    \includegraphics[width=.8\textwidth]{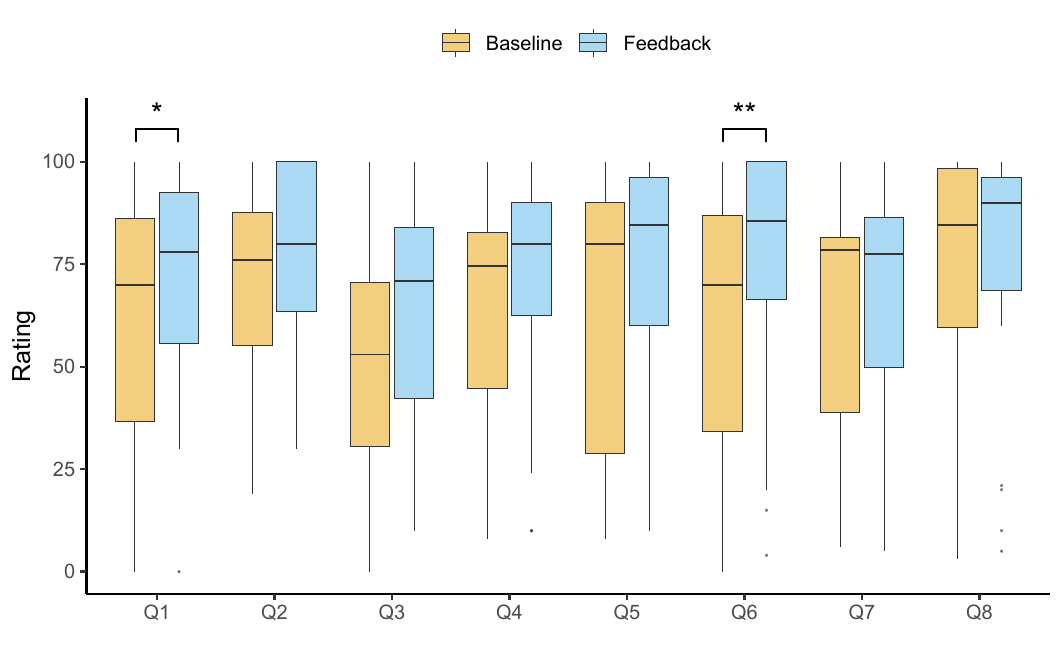}
    \caption{Ratings for our custom questionnaire (see~\Cref{tab:custom_questions}). Q3 ("The system helped me to gain new insights into my timing while dancing."), and Q6 ("I used the system to find back into the rhythm.") show significant differences for the conditions (\fb{} vs. \base). Significant questions are marked with *.}
    \Description{Ratings for our custom questionnaire (see Table 1). Q1 ("I noticed the system while dancing."), and Q6 ("I used the system to find back into the rhythm.") show significant differences between the feedback and the baseline condition. Significant questions are marked with * (Q1) and ** (Q6).}    
    \label{fig:results_custom}
\end{figure}

In addition to the objective performance metrics, we analyzed questionnaire responses assessing participants’ subjective experiences with \prototype.
\paragraph{Custom Questions}
An overview about the outcomes of our customized questions (listed in \Cref{tab:custom_questions}) is depicted in \Cref{fig:results_custom}. 
Our statistical analysis revealed a significant main effect of condition for Q1: ``I noticed the system while dancing.'' ($\beta = 28.73$, $\stderror = 13.22$, $t(22.0) = 2.17$, $p = .041$), with higher scores for the \fb{} condition, indicating that the additional clapping feedback during the \fb{} condition was actively perceived by participants. Likewise, for Q6: ``I used the system to find back into the rhythm.'', we found a significant main effect of condition ($\beta = 30.23$, $\stderror = 9.17$, $t(22.0) = 3.30$, $p = .003$), with higher ratings for the \fb{} condition, indicating that participants made use of the clapping feedback to recover the rhythm. We found no further significant main effects, nor any significant interaction effects.
Descriptive analysis revealed, that \fb{} scored higher than \base{} for all questions except for Q7: ``The system was in line with my body perception.'', which in part might indicate that participants struggled to align their own bodily perception of the rhythm with the correct rhythm as dictated by the system. The remaining items addressed broader aspects, such as general helpfulness (Q2), timing awareness (Q3, Q4), and overall perceived support from the system (Q5, Q8). Higher ratings for \fb{} indicated that the clapping feedback was generally perceived positively. 

\paragraph{NASA TLX}
We used the NASA TLX~\cite{hart1988NASATLX} to measure the perceived task load. Participants rated the load similar for \fb{} ($\samplemean=55.6, \samplesd=14.5$) and \base{} ($\samplemean=58.1, \samplesd=17.8$). We found a significant negative main effect of dancing experience ($\beta = -0.629$, $\stderror = 0.167$, $t(36.55) = -3.77$, $p = .001$). Thus, as anticipated, TLX scores decreased significantly with higher dancing experience, i.d. perceived workload was lower for more experienced dancers. No effects of condition or interaction were found.

\paragraph{Self-Appraisal}
In terms of the self-appraised performance score~\cite{mamassis2004effectsmental} the participants rated their performance generally higher in the \fb{} condition with $\samplemean=28.4 $ ($\samplesd=5.04 $) than in the \base{} condition with $\samplemean=27.4 $ ($\samplesd=4.3$). We found a significant positive main effect of dancing experience ($\beta = 0.118$, $\stderror = 0.056$, $t(31.57) = 2.12$, $p = .042$), indicating that self-appraisal scores increased with higher dancing experience.

\paragraph{Perceived Creepiness of Technology Scale}
The \fb{} received slightly higher ratings on the perceived creepiness scale~\cite{woźniakCreepyTechnologyWhat2021} (\fb: $\samplemean=20.5, \samplesd=8.4$ vs. \base: $\samplemean=19.0, \samplesd=6.7$, 7-point Likert, higher values indicate higher level of creepiness). Neither the main effects of condition or dancing experience, nor their interaction, were significant.

\paragraph{Flow Experience Scale}
Statistical analysis also did not report any significant effects on the flow scale~\cite{rheinbergErfassungFlowErlebens2003} for \fb{} (Flow: $\samplemean=4.71, \samplesd=1.01$; Worry (low is better): $\samplemean=3.21, \samplesd=0.95$) vs. \base{} (Flow: $\samplemean=4.65, \samplesd=1.04$; Worry (low is better): $\samplemean=3.26, \samplesd=0.96$).

\subsubsection{Interviews}
\label{subsec:s_results_interviews}
Through the thematic analysis, we identified the following themes under which we grouped the statements: 

\begin{description}
    \item [Body Proprioception and Perception of Rhythm:] The ability to align bodily movements with the time constraints dictated by the music. This also includes that the participants know where they are in the song and know how to align the movements with it.
    \item [Perception of the Feedback Provided by Skeleton Dance:] That the feedback was noticeable, in line with the perception, and was not disturbing. How the participants perceived it and used it. 
    \item [Skeleton Dance Supported Rhythm:] The system supported the participants in understanding the rhythm and getting back into it after losing it.
    \item [Suggested feedback modalities:] Potential other or additional feedback options.
    \item [Cognitive effort:] Cognitive effort required during the dancing and regarding the system experience.
    \item [Usability:] Using and interacting with the system.
    \item [Usage suggestions:] Potential target groups and usage contexts. 
\end{description}

\paragraph{Body Proprioception and Perception of Rhythm}
The participants often commented on their body proprioception and rhythm perception. 
As novices, they reported difficulties synchronizing movements with music and finding the song’s rhythm in the first part~(P15). However, some noted that they did not perceive the movements as too tricky without music, but combining them with rhythm increased difficulty.
Once music was introduced, executing the dance sequence became more challenging for some participants, showcasing the difficulties in aligning the bodily and temporal demands~(P3).

In that sense, they also struggled with identifying the duration of the movements in the dance sequence with the music but used the video to analyze speed and duration. 
\begin{quoting}
    The third step, [...] where you’re doing the knee thing [...]. I didn’t understand the speed of it the first few times [...], and I was slower than the beat, so I kind of missed a few beats there [...]. When I looked at your video, I realized how you’re doing it, what the speed of it is [...]. After a couple of times, I was able to get it right.~(P19)
\end{quoting}
Participants generally reported that they perceived the second round as easier than the first one, making them feel more confident even without the clapping feedback~(P20, P25). 

Further, participants used the feedback to calibrate their own perceptions. With the help of the feedback, they expect feeling reassured that their performance was better than expected.
\begin{quoting}
    I think if the system additionally says, ``Hey, you’re not actually doing that much wrong,'' then I think it would be [...] best because [for] me, the first time, I thought [...], maybe I’m not doing everything so wrong, but it just seems that way to me.~(P3)
\end{quoting}
Conversely, when the clapping feedback started, it showed them a mismatch between perceived and actual rhythm performance, making them opt for for a cleaner and more precise synchronization~(P22).

\paragraph{Perception of the Feedback Provided by Skeleton Dance}
Next, we asked the participants how they perceived and utilized the feedback provided by \prototype{}. Participants found that the feedback supported their sense of rhythm, which sometimes happened subconsciously, as the enhanced beat inherently supports rhythm. Further, they used the system to find the rhythm and recognize their own rhythm errors.

The safety and trust finding is supported by statements from participants who received clapping feedback in the first round of the study. They mentioned they missed it during the second round, where they did not get auditory feedback.
However, some found the connection between feedback and their movement unclear, perceiving it as random or difficult to interpret, blaming it on their poor sense rhythm (P7).
While a few participants did not notice or detect a clear pattern in the clapping feedback (P20), others perceived it as seamlessly integrated into the music and familiar from prior learning contexts.
\begin{quoting}
    $[$Understanding the clapping feedback] was easy. Since it was exactly in harmony with the music and also this normal beat that you know from music lessons or something. [...]~(P25)
\end{quoting}
Others reported not consciously using the clapping feedback per se. Still, when reflecting on it afterward, they believed it subconsciously influenced their performance, timing and rhythm awareness~(P10). 
Even if they did not think about it actively while dancing, participants perceived the mirror as a helpful complementary feedback channel~(P4).

\paragraph{Skeleton Dance Supported Rhythm}
Moreover, most participants found that the \prototype{} system supports rhythm. 
The feedback generally helped them familiarize themselves with the beat, enhancing their perception of dance performance, especially when the clapping feedback was present in the first round.
\begin{quoting}
    In the beginning, with assistance, I heard the beat while practicing, and it somehow helped me. I thought it was my own clapping at the first time. When I was putting my step on the floor, I felt like the beats were synchronous with my steps, and I felt more confident when I was stepping.~(P20)
\end{quoting}
Still, some missed the clapping feedback in the second round if they had experienced it earlier~(P9).

The clapping feedback also helped them find the rhythm in the beginning~(P12) or to find back upon losing it. It enabled them to focus on the music's beats and figure out when to restart dancing.
\begin{quoting}
    That made this beat more obvious. In case you lose the rhythm, it's very easy to get back into it and follow the beat because of that feedback. [...] I know where to start.~(P19)
\end{quoting}
The clapping sound shifted their attention toward the beat and their movement~(P9). 

The participants also applied a common approach themselves that teachers use in dance classes (cf.~\Cref{sec:e_interviews}). They silently counted the beats. The video tutorials included counting the beats as well. It is a valuable tool to help beginner dancers who struggle with synchronizing movement to rhythm.
\begin{quoting}
   I also had to keep counting myself the whole time; otherwise, I don't think it would have worked at all.~(P8)
\end{quoting}

\paragraph{Suggested Feedback Modalities}
They suggested more options or variations for the auditory feedback, such as an additional cues marking the start of a new dance movement or the start of a new repetition. 

In addition to the existing clapping and mirror feedback, participants envisioned the potential benefits of other visual cues, such as lights flickering in sync with the music or a ghost dancer helping perform the sequence in rhythm. They further suggested color-coded feedback, e.g., limbs or lights, to indicate whether a leg is (not) lifted high enough~(P17, P18).

Many beginner dancers wanted more explicit feedback, indicating whether they were doing well or badly. 
\begin{quoting}
    $[$If$]$ I'm actually on the beat, I would have been happy if I would have had a thumbs up or [something] like, "You are doing great. You're [in/out of] the rhythm,[...]" [it] would have helped more.~(P24)
\end{quoting}

They also commented on the study procedure and appreciated that the experimenter left the room and did not observe them during their practice. Beginners, in particular, quickly feel overwhelmed and intimidated, which may keep them from dancing and trying it.
\begin{quoting}
    $[$As$]$ I said, I can’t dance at all. I’m also a bit uncomfortable with that [...] It was also good that I was alone in this room. I think it would have been bad if you had been here; I would have been ashamed.~(P7)
\end{quoting}
\paragraph{Cognitive Effort}
Given that they were all beginner dancers, the process of acquiring and retaining dance sequences while synchronizing them with music was unfamiliar to them, requiring a higher cognitive effort compared to what experienced dancers would have encountered. They had to focus more on the music and could not pay that much attention to it~(P4).
Participants often missed beats while trying to remember the next movement, highlighting the challenge of balancing memory and rhythm.
\begin{quoting}
    $[$I feel like$]$ it's best if you just intuitively trust your muscle memory because as soon as you start thinking, you get out of [rhythm] right away.~(P22)
\end{quoting}
However, the subtle clapping feedback reduced the required cognitive effort by blending into the background while simplifying beat perception, making it easier to recognize rhythm during practice~(P26).

\paragraph{Usability}
Moreover, we discussed the usability of \prototype{}. Most participants stated that \prototype{} made dance practice fun, and they felt supported. Further, the system was non-intrusive and easy to use, allowing flexible practice when and wherever they want~(P14). 

Participants also found the chosen dance sequences engaging and suitable for novices, noting that the instructions were presented clearly and well-structured, which facilitated step-by-step learning. 
\begin{quoting}
    The video itself is clear; everything is well structured. [A]fter each movement a repetition. I was able to train one movement first, then all together.~(P13)
\end{quoting}
However, some participants voiced that they, as beginners, need more support than the current system offers, particularly on how to combine movement with music and how to maintain rhythm~(P4). 

\paragraph{Usage Suggestions}
Lastly, we discussed potential usages of \prototype{}, primarily envisioning it as a tool for remote training. 
This setting lowers barriers for inexperienced users, allowing them to practice without the pressure of a studio environment.
In addition, the system provided feedback similar to what a dance teacher would offer.
\begin{quoting}
    Definitely at home. I often hear people say, a) I don't go to this [...] sport because there are only professionals, and I want to have a certain basic level first, and then I would [...] go there. I know that about myself, too [...]. Therefore, as home training [it gives] people the opportunity to get into dance.~(P5)
\end{quoting}
Participants also recommended the system for additional remote practice, for example, when beginners struggle to keep up in dance class, they can practice it at home first while still receiving feedback~(P14). 
They emphasize that feedback is especially important for beginners, as their rhythm perception is still developing.
\begin{quoting}
    $[$A$]$ lot of people that don't have the intuition for rhythm. [Some] people [...] just don't have it. It's [...] difficult for them. I think for those kind of people, this would be really, really helpful when you have the feedback [...]. [T]hen you know exactly where the beat is.~(P19)
\end{quoting}


\section{Discussion}
\label{sec:discussion}
Our investigation showed that beginner dancers can gain a better understanding of rhythm with the appropriate feedback. We could observe from the interview answers that they genuinely developed a feeling for the rhythm while having the feedback as a safety net. In the following, we discuss whether the design requirements formulated in \Cref{subsec:designPT} (\textbf{D1-5}) are met and highlight implications of our findings for future systems.


\subsection{\prototype{}'s Clapping Feedback Supports Rhythm During Practice}
Participants consistently reported that the feedback was noticeable and available when needed, as reflected in the questionnaire responses (cf. Q1, \Cref{subsec:s_procedure}). They further indicated that the feedback helped them find back into the rhythm after losing it (cf. Q6, \Cref{subsec:s_procedure}). Participants described feeling safer while practicing when the clapping overlay was present, as it made the beat of the music more prominent.
Statements from the interviews further showed that they actively used the clapping feedback when it was available in that condition to get back into the rhythm upon losing it. \prototype's immediate feedback (\textbf{D1}) helps to identify errors to ensure that they do not creep in. 

\subsection{Dancing Experience Moderates the Usefulness of Feedback}
The objective metrics showed that prior dancing experience played a central role in rhythmic performance across conditions. For \emph{\beatshit{}}, feedback was most beneficial for less experienced dancers, while more experienced dancers demonstrated higher rhythmic accuracy in the \base{} condition, indicating that more experienced dancers were better able to maintain rhythm without support. However, this relationship was moderated by condition, resulting in a cross-over pattern in which the advantage of feedback diminished as dancing experience increased. This interaction suggests that while clapping feedback supported less experienced dancers, it may have interfered with or become redundant for more experienced participants.
A similar pattern emerged for \emph{\amtlost{}}, where higher dancing experience was associated with fewer rhythm losses overall, indicating increased rhythmic stability with experience. These findings were in line with our supporting metric \emph{\fbduration{}}, as the clapping feedback was active for longer for participants with less dancing experience during their practice sessions.
\emph{\meanofftimes{}} showed no significant effects. However, descriptive analysis indicated that the \fb{} condition supported participants in recovering faster once rhythm was lost.

Subjective measures, as captured through the NASA TLX and the self-appraisal questionnaires, align with the patterns observed in the metrics on rhythm accuracy. Here, participants with greater dancing experience reported significantly lower perceived workload and higher self-appraisal, suggesting that experienced dancers felt more confident and less strained during the task.

These findings indicate that \prototype{}’s feedback was most beneficial for novice dancers, while its utility decreases as dancers gain experience. As such, dancing experience emerges as a strong moderating factor, particularly for lightweight feedback designs.

\subsection{Motor Learning Theory as a Design Lens}\label{subsec:mlt_design_lens}
Although our work did not aim to empirically evaluate motor learning theories (cf.~\Cref{sec:method}), these theories provide a useful interpretive lens for reflecting on our design implications derived in \Cref{sec:designimpl}. Participants’ reports of increased rhythmic awareness, confidence, and the ability to recover from rhythm errors resonate with motor learning theories that emphasize the role of timely, externally focused feedback in early stages of skill acquisition. From this perspective, our design choices, including immediate feedback (\textbf{D1}), an external focus on rhythm through clapping (\textbf{D2}), a deliberate focus on a single aspect of dance practice (\textbf{D3}), feedback that adapts to the learner’s momentary needs (\textbf{D4}), and a lightweight setup that supports accessible, self-directed practice (\textbf{D5}), can be understood as supportive practice conditions that help learners notice and respond to timing deviations. Our design principles describe how \prototype{} structures early-stage practice in a way that aligns with theoretical accounts of the gradual formation and refinement of generalized motor programs through repeated engagement.

\subsection{Limitations}
In this study, we evaluated \prototype{} with only two dance sequences. The two sequences can be considered exemplary for many other dance sequences and were approved by dance teachers. In theory and according to motor learning literature, once the feeling for the rhythm and dancing in sync with the music is established as a \GMP, it should be transferable to other dances and dance styles with little effort~\cite{schmidtMotorControlLearning2018}.

Further, we conducted the study over a limited time span, so we could not ascertain quantitative improvement. A longitudinal evaluation would be necessary to capture changes in dance expertise over time and to model learning trajectories during practice. Such an approach could further examine the consolidation of \GMPs{} and more directly link the use of \prototype{} to motor learning theory. In our work, the qualitative feedback already shows that our approach is promising. The participants reported that they actively used the clapping feedback and felt safer and more confident with this support.

Some participants remarked that they would have liked to get more explicit feedback, e.g., encouraging or reassuring positive feedback through visual highlights. In \prototype, we deliberately chose this minimal feedback design for two reasons: (1) our feedback method is based on well-established approaches in real-world dance classes, and (2) prior research indicates that excessive feedback can potentially impede practice~\cite{gibbons2013feedbackdance}.

Likewise, \prototype{} is deliberately designed for novice dancers. By offering only basic, rhythm-focused feedback, the system prioritizes early engagement and accessibility rather than long-term mastery. Although its usefulness decreases as dancers gain expertise, the minimal setup and broad applicability highlight the value of lightweight feedback systems in supporting initial practicing phases and lowering barriers to entry.

\section{Conclusion}
In this study, we explored how an interactive feedback system, called \prototype{}, can support novice dancers in practicing dance by providing rhythm feedback. We conducted expert interviews with dance teachers to identify common difficulties faced by beginner dancers and consulted with them to determine the most effective feedback approach, linking our design and teaching approach with established motor learning theories. 
The implemented algorithm generated human-like clapping sounds that accentuated the music's beats. It was added as auditory cues synchronized with the music, aiding users in regaining rhythm when lost. 
In our user study, participants reported feeling better supported. It significantly helped them in finding back into the rhythm after losing it. The clapping feedback took away some strain by taking care of the rhythm.
They found the system user-friendly and suitable for remote use, benefiting individuals seeking independent practice. Our findings emphasize the value of interactive beat-enhancing feedback for beginner dancers, fostering motivation and practice. Customizable feedback options and the use of a basic webcam for data collection further highlight the system's potential in dance education.

\bibliographystyle{ACM-Reference-Format}
\bibliography{skeletondance, pawel}
 
\appendix
\section{Semi-Structured Interview: Dance teacher}
\label{sec:interviewprotocol}
Introduction of the interviewer, details about procedure, informed consent, goal of the project. Asking if there are any questions before starting the interview.\\
\textbf{Demographics:}
\begin{itemize}
    \item How old are you?
    \item Which pronouns do you claim for yourself? He/him, she/her, they/them?
    \item What is your profession? In what field do you work?
    \begin{itemize}
        \item Do you work full-time as a dance teacher?
    \end{itemize}
\end{itemize}

\textbf{Dancing and Teaching:}
\begin{itemize}
    \item How long have you been dancing?
    \item How long have you been a dance teacher?
    \item How did you come to teach dance?
    \item What style do you teach?
    \begin{itemize}
        \item What do you particularly like about this style?
    \end{itemize}
    \item Is there a dance/direction that you particularly enjoy dancing yourself, or teaching? If so, why?
    \begin{itemize}
        \item Can you describe the characteristics / typical movements for this dance style? (Fast/slow movements, a lot of leg/arm movement, synchronization)
    \end{itemize}
    \item Do you remember some of your own dance lessons? How were they structured?
    \item What do you think are the most important components in dance?
    \begin{itemize}
        \item Timing
        \item Correct technique
        \item Enjoyment
        \item Good body position
        \item Is something missing? $\rightarrow$ What?
        \item Please sort the components by importance -- for you personally
        \item Follow-up questions:
        \begin{itemize}
            \item Why have you sorted it exactly this way?
            \item Why is timing more important to you than technique, for example?
            \item \ldots
        \end{itemize}
    \end{itemize}
    \item What makes a good dancer? What are his qualities?
\end{itemize}

\textbf{Scenarios:}
\begin{itemize}
    \item You teach a dance group a completely new dance sequence, possibly including complete beginners.
    \begin{itemize}
        \item What is your approach for teaching it to the students?
        \item How is a lesson structured?
        \item What do you emphasize most?
        \item Do you pay special attention to something?
        \item How do you make sure that body position is maintained?
        \item How do you make sure that timing is maintained?
    \end{itemize}

    \item If you later realize that the motor movements have been implemented quite well, but students still have problems with the timing to the music.
    \begin{itemize}
        \item How do you deal with this situation?
        \item How do you deal with it when there is only one pupil, for example? So not several pupils?
        \item Do you have any special tricks?
        \item Is this important to you at all?
    \end{itemize}

    \item What generally helps pupils to keep time? Or to recognize it in the first place? How can they learn this? To what extent do you provide support?

    \item Is ``lack of rhythm'' a problem you encounter more often in your lessons?

    \item What do you do if you notice that even after a few lessons there is still the same/a rhythm problem?

    \item Which methods do you normally use in class? What resources might be useful for you in teaching dance. Is there anything that could assist you or that would be helpful for you to teach a dance?
    \begin{itemize}
        \item Negatives about those tools?
        \item Do you use metaphors?
        \item Do you conceptualize the movement, if yes, how? Or do you just actually think about the movement itself?
        \item Have you ever had to deal with technical systems that helped or supported training in any sport?
    \end{itemize}

    \item Have you already taught a virtual/digital dance class?
    \begin{itemize}
        \item What tools did you use for this?
        \item Do you think you explain dances/figures differently in a digital dance class? If so, can you give me an example?
        \item What do you miss about teaching in a digital context compared to non-digital classes?
        \item Do you have information on what are the main problems students face when taking a virtual dance class/unit?
        \item Negatives about those tools?
    \end{itemize}

    \item Could you imagine that people could be taught certain parts, which is now more about technique, basics, via some kind of aid?
    \begin{itemize}
        \item If it's about basic movement sequences, what would a system have to be able to do? How would it have to teach that?
    \end{itemize}

    \item What makes a (good) trainer?
\end{itemize}

\textbf{Support with technology:}
\begin{itemize}
    \item Can you think of ways how technology could support students in learning dance?
    \begin{itemize}
        \item In class?
        \item At home? $\rightarrow$ Student by himself/herself
    \end{itemize}

    \item Can you think of ways how technology could support students in learning timings?
    \begin{itemize}
        \item In class?
        \item At home? $\rightarrow$ Student by himself/herself
    \end{itemize}

    \item What different sensors could be used?

    \item How could the feedback be provided to the user?
    \begin{itemize}
        \item Haptic / visual / auditory
    \end{itemize}

    \item Would you use such a system? What would a system need to have, in order for you to use it?
    \begin{itemize}
        \item In which area would you most likely use such a system?
        \item For which target group?
    \end{itemize}

    \item What should be main considerations when building such a system? (Too much blinking, vibrating, \ldots)

    \item It makes it easier for people to think about e.g.\ feedback, if I say: imagine having a system that knows where the user is and how the body position is at any given time $\rightarrow$ what feedback would you want? How would you want that?
    \begin{itemize}
        \item Posthoc vs.\ immediate?
        \item When would feedback best need to be given?
    \end{itemize}

    \item Imagine: In terms of technology, anything is possible.
    \begin{itemize}
        \item Do you think a training effect can be achieved through such a system?
        \item Do you see any problems?
    \end{itemize}
\end{itemize}

\textbf{Interview closing:} Anything else you want to add? Do you have any questions? -- Thanks and goodbye
\end{document}